\documentclass[12pt]{article}
\usepackage{amsmath}
\usepackage{amsfonts}
\usepackage{amssymb}
\usepackage{appendix}

\begin{document}
\title{WEYL TIME-EVOLUTION-OPERATOR IN PLANAR BIANCHI-TYPE-I UNIVERSES}
\author{MATTHIAS WOLLENSAK\footnote{matthias.wollensak@uni-jena.de} \\Theoretisch-Physikalisches Institut,
\\Friedrich-Schiller-Universit$\ddot{a}$t Jena,
\\Max-Wien-Platz 1, D-07743 Jena, Germany}

\maketitle

\begin{abstract}
In this paper, Dirac`s  equation in anisotropic Bianchi-type-I background spacetimes is treated w.r.t. orthonormal frames. Specializing to the massless spinor case and metrics with power law scale factors and planar symmetry, an analytical expression of the approximate time-evolution-operator is derived. By use of this operator all approximate spinor solutions can be generated. By construction, these solutions agree asymptotically and at early times with the exact solutions.

The operator approach is also well adapted for studying the case of small deviations from conformally flat backgrounds. In particular, it can be shown that the limiting case of vanishing anisotropy renders the correct result.
\end{abstract}

\section*{1. Introduction}
The behavior of particles in curved backgrounds obeying the Dirac equation has long been of considerable interest in cosmology and astrophysics. An early treatment of the massless case has been given in the pioneering work of Brill and Wheeler \cite{Brill}, the quantized massive spin-$\frac{1}{2}$ field in (conformally flat) Friedmann-Lemaitre-Robertson-Walker (FLRW) background spacetimes was first studied by Parker \cite{Parker}. Further questions in connection with the quantization of fermionic fields in FLRW and conformally flat FLRW (fFLRW) backgrounds have been addressed by several authors \cite{FLRW-Quant}.

At least in a few cases, it is relatively easy to find exact classical solutions of Dirac`s equation in isotropic backgrounds. For example, Barut and Duru investigated massless and massive fermions in fFLRW spacetimes with power law expansion, and in the steady-state part of de Sitter spacetime \cite{Barut}. The latter model had also been studied by Cotaescu, who in addition performed the quantization of the spin-$\frac{1}{2}$ field \cite{Cotaescu}, and by Candelas and Raine using a path integral approach \cite{Candelas1}. Fermion propagators in fFLRW backgrounds with constant deceleration have been treated by Koksma and Prokopec \cite{Koksma}. 

However, if one is occupied with questions concerning the early universe, then one inevitably encounters the problem of anisotropy. Although the present day universe seems to be isotropic to a very high degree, this was not necessarily the case in an early phase of development of the universe. For example, decades ago, Zel`dovich pointed out that if one starts with an anisotropically expanding Kasner universe, which belongs to a subtype of Bianchi-type-I (BI) universes, and takes into account quantum effects in the vicinity of the initial singularity, this could lead to an isotropization of the universe at the Planck time-scale due to particle creation processes \cite{Zel`dovich2}. A semiclassical calculation by Hu and Parker lent further credit to this idea \cite{Hu}. They investigated a quantized massless conformal scalar field in a planar BI (pBI) universe. A further motivation for the interest in Kasner universes as background spacetimes is given by the work of Belinskii et al. \cite{BKL}, and Misner \cite{Misner}, who realized that Bianchi-type-IX universes can be described by sequences of Kasner epochs, if one moves backwards in time towards the initial singularity.
More recently, the issue of anisotropy and in particular of BI and Kasner spacetimes has been discussed in the context of preinflationary scenarios of the universe  \cite{Pitrou}, \cite{Gm}, and also in vector inflation \cite{Armendariz} and gauge inflation models \cite{Malaknejad}.

Classical $3$-space independent solutions of Dirac's equation in BI spacetimes have been investigated e.g. by Henneaux \cite{Henneaux}, and by Saha and Boyadjiev \cite{Saha}. To obtain analytical results when treating quantized fermions propagating in anisotropic backgrounds, one usually resorts to a  perturbative treatment of the spacetime. For example, in a method utilized by Zel`dovich and Starobinsky, small anisotropic perturbations about a fFLRW spacetime are considered \cite{Zel`dovich1}. For a model with a special form of  weak anisotropy, Birrell and Davies performed the quantization of the massive scalar field \cite{Birrell}, and Lotze treated the quantization of massive fermions in this background \cite{Lotze}.

However, the fact remains that usually no exact (even massless) classical spinor solutions in BI (even pBI) backgrounds are at disposal. Nevertheless, one can try to calculate solutions valid at late and early times, resp. But it is not clear how (and whether at all) the short-time and asymptotic solutions found in this way match in the sense that to a given asymptotic solution the corresponding early-time solution can be uniquely related (and vice versa). This suggests to tackle the problem by trying to determine $approximate$ solutions of the Weyl-Dirac equation in BI backgrounds. These solutions are as close as possible to the usually not available exact ones, since by definition they match them at early and late times. Hence, they automatically solve the initial value problem, which significantly limits the physical value of purely asymptotic solutions. The approximate Weyl time-evolution-operator (TEO) derived in this work provides these approximate solutions.

The paper is organized as follows: In sec. 2, we set up Dirac`s equation in BI backgrounds. In sec. 3, we specialize to the case of massless fermions in pBI spacetimes with power-law scale factors. For those backgrounds, an exact expression of the Weyl TEO is found. This result can be used to derive a simple parameter transformation (PT) generating all solutions of a given equivalence class of exact (or approximate) massless spinor mode solutions in pBI backgrounds, provided one knows a single arbitrary solution of this class \cite{Wollensak1}. It is shown that this PT is equivalent to a constrained diffeomorphism. We then explicitly determine an analytic expression for the approximate Weyl TEO, which is nonperturbative w.r.t. the underlying background spacetime. The outcome of this novel approach is compared in sec.s 4.1 - 3  with the solutions of exactly solvable models, and in sec. 4.4 with a model whose background is described by Kasner's anisotropic planar solution. In this case no exact spinor solutions are known. In sec. 5 we investigate the Weyl TEO in anisotropically perturbed conformally flat backgrounds, and in sec. 6 some general properties of the TEO are discussed.

Throughout this work we use $c = 1$, where $c$ is the speed of light.
\\
\section*{2. Orthonormal Frames and Dirac's Equation}
In a coordinate frame with cosmic time $t$, the line element of a BI universe reads

\begin{equation}
\label{Gl1}
ds^2 = dt^2 - \sum\limits_{i=1}^{3} 
\alpha_{i}^{2} (t)(dx^{i})^2   ,
\end{equation}
with metric tensor $\boldsymbol{g} = g_{\mu\nu} dx^{\mu} \otimes dx^{\nu}$, where $g_{\mu\nu} = \mathrm{diag}(1,- \alpha_{1}^{2},- \alpha_{2}^{2},- \alpha_{3}^{2})$. A natural choice of an orthonormal frame at $p \in M$ is given by the covectorfields  $\Theta^0 = dx^0 \equiv dt,\ \Theta^j = \alpha_j(t) \, dx^j$ (no sum), which constitute bases of the fibers $T^{\ast}_p(M)$ of the cotangent bundlespace $\mathcal{B}(M, T^{\ast}_p(M)) \equiv T^{\ast}(M) := \bigcup_{p \in M} T^{\ast}_p(M)$. The base space $M$ denotes a differentiable pseudo-Riemannian manifold endowed with metric $\boldsymbol{g}$. Likewise, $e_0 = \partial_0 \equiv \partial_t, \  e_j = \alpha_j^ {-1}(t) \, \partial_j$ (no sum) are the corresponding basis vectorfields of the fibers of the tangent bundlespace $\mathcal{B}(M, T_p(M)) \equiv T(M) := \bigcup_{p \in M} T_p(M)$. The line element (1) is then given by $ds^2 = (\Theta^0)^2 - (\Theta^1)^2 - (\Theta^2)^2 - (\Theta^3)^2$, and the metric tensor assumes the form $\boldsymbol{g} = \eta_{\mu\nu} \Theta^{\mu} \otimes \Theta^{\nu}$ with $\eta_{\mu\nu} = \boldsymbol{g}(e_{\mu}, e_{\nu} )  \equiv \mathrm{diag}(+1, -1,-1,-1)$,  since $\Theta^{\mu}(e_{\nu})  = \delta^{\mu}_{\nu}$.  The exterior derivative of $\Theta^\mu$ reads: $d\Theta^{\mu} = - C^{\mu}_ {\ \alpha \beta} \Theta^{\alpha} \wedge \Theta^{\beta}$, with three nonvanishing commutation coefficients ($C^{\mu}_ {\ \alpha \beta}  =  - C^{\mu}_ {\ \beta \alpha}$):

\begin{equation}
\label{Gl7}
C^j_ {\ 0j} = - \dot{\alpha_j}/\alpha_j   ,
\end{equation}
($j = 1, 2, 3$, no summation). Correspondingly, the commutator of the vectorfields is defined to be: $[e_{\mu}, e_{\nu}] = C^{\alpha}_ {\ \mu \nu} e_{\alpha}$.
Only in bundlespaces where the fibers are given by the tangent spaces $T_p(M)$ of the underlying base manifold $M$, one can define a soldering form $\mathcal{X} = e_{\nu} \otimes \Theta^{\nu}$ \cite{Trautman}, \cite{Sexl}. This soldering form can be viewed as a vector-valued 1-form, and applying the exterior covariant derivative to $\mathcal{X}$ results in $D \wedge \mathcal{X} = e_{\nu} \otimes T^{\nu}$, where $T^{\nu}  = d\Theta^{\nu}  + \omega^{\nu}_ {\ \mu} \wedge \Theta^{\mu}$ denotes the torsion 2-form. The exterior covariant derivative of the soldering form must be distinguished from its  covariant derivative, which satisfies: $D \, \mathcal{X} \equiv 0$. To derive Dirac`s equation in curved spacetime, we use as ansatz for the connection 1-form:

\begin{equation}
\label{Gl8}
\omega^{\nu}_ {\ \mu}  = L^{\nu}_ {\ \mu \alpha}\Theta^{\alpha}   .
\end{equation}
Because we are dealing with a pseudo-Riemannian manifold, the two conditions metricity: $dg_{\mu \nu} = \omega_{\mu \nu} + \omega_{\nu \mu}$, and vanishing torsion can be utilized to determine $\omega_{\mu \nu}$: $T^{\nu} = 0$ implies: $C^{\mu}_ {\ \alpha \beta} = L^{\mu}_ {\ \beta \alpha} - L^{\mu}_ {\ \alpha \beta}$, and from metricity follows: $\omega_{\mu \nu} = -\omega_{\nu \mu}$ and $L_ {\alpha \beta \gamma} = - L_ {\beta \alpha \gamma}$, since $g_{\mu \nu} = \eta_{\mu \nu}$ in an orthonormal frame. Thus, $C^j_ {\ 0j} = - L^j_ {\ 0j} = L_ {j0j}$ (no sum), and one finds for the nonzero entries of the connection 1-form:

\begin{equation}
\label{Gl11}
\omega^j_ {\ 0} = - C^j_ {\ 0j} \Theta^j \equiv \omega^0_ {\ j}.
\end{equation}
The covariant differentiation of a Dirac bispinor is given by the 1-form

\begin{equation}
\label{Gl12}
(D\psi)^{I} = d\psi^{I} + \frac{1}{8} \ \omega^{\mu}_ {\ \nu} \ [\gamma_{\mu}, \gamma^{\nu}]^{I}_{\ L} \psi^{L}
\end{equation}
(with spinor indices $I, L = 1, 2$) \cite{Sexl}, where the $\gamma^{\mu}$ are flat space Dirac matrices in standard representation: $\gamma^0 = i \sigma^3 \otimes  \sigma^0, \ \ \gamma^j = i \sigma^2 \otimes  \sigma^j$. From ($ \ref{Gl12}  $) follows $[D, \gamma_5] = 0$, where $\gamma_5 := i  \gamma^0 \gamma^1 \gamma^2 \gamma^3$. Hence, in the massless case the chirality operator $\gamma_5$ generates a global symmetry of the free Lagrangian. In spacetime (\ref{Gl1}) one gets owing to (\ref{Gl11}) and (\ref{Gl12})

\begin{equation}
\label{Gl13}
\gamma^{\mu}  (D \psi) (e_\mu) \equiv \gamma^\mu D_{e_\mu} \psi = \left\lbrace \gamma^\mu e_\mu  - \frac{1}{4} \ \sum\limits_{j=1}^{3} \ C^j_ {\ 0j} \gamma_j [\gamma_0, \gamma_j] \right\rbrace  \psi   ,
\end{equation}
so that Dirac`s equation takes the form\footnote{For the analogous treatment of Maxwell`s equations see ref. \cite{Wollensak2}.}

\begin{equation}
\label{Gl14}
\left\lbrace e_0 - \gamma_0 \sum\limits_{j=1}^{3}
\gamma_j e_j  +  {\cal{C}} + i \gamma_0 m \right\rbrace \psi = 0   ,
\end{equation}
where ${\cal{C}} := - \frac{1}{2} \sum^3_{j=1} C^j_ {\ 0j} = \partial_t \, \ln(|g|^{1/4})$ and $g$ = det $g_{\mu \nu}$. Spatial translation invariance of the line element ($ \ref{Gl1}  $) motivates the ansatz

\begin{equation}
\label{Gl15}
\psi_\textbf{k}(\textbf{x},t) = c_\textbf{k} \, e^{i\textbf{k} \textbf{x}} \left( \begin{array}{c} \varphi(\textbf{k}, t)  \\ \chi(\textbf{k}, t)  \end{array} \right) 
\end{equation}
with normalization constant $c_\textbf{k}$, and $\varphi, \chi$ denote Weyl-spinors. Eq. ($ \ref{Gl14}  $) can then be rearranged as a coupled system for those spinors:

\begin{equation}
\label{Gl16}
\begin{aligned}
e_0 \left(e^{+ imt} |g|^{1/4 } \varphi \right) - i \textbf{p} \boldsymbol\sigma \, e^{+ imt} |g|^{ 1/4 } \chi & = 0   ,
\\
e_0 \left(e^{- imt} |g|^{ 1/4 } \chi \right) - i \textbf{p} \boldsymbol\sigma \, e^{- imt} |g|^{ 1/4} \varphi  &= 0    ,
\end{aligned}
\end{equation}
where the physical 3-momentum $\textbf{p}$ is defined by

\begin{equation}
\label{Gl17}
p_j(t) = k_j/\alpha_j(t) .
\end{equation}
In the massless case the system (\ref{Gl16}) decouples with ansatz $\chi = \mp \varphi$. With (\ref{Gl15}) the chirality eigenstates read:  $\psi_\textbf{k}^{(\mp)} = c_\textbf{k}^{(\mp)} e^{i \textbf{k}  \textbf{x}} (\varphi^{(\mp)}, \mp \varphi^{(\mp)})^T$. For convenience, we introduce the spinors $\phi^{(\mp)} =   (\phi_1^{( \mp)},  \, \phi_2^{(\mp)})^T$, related to $\varphi^{( \mp)}$ by:

\begin{eqnarray}
\label{Gl18}
\phi_L^{(\mp)}(\textbf{k}, t) \, = \,  |g(t)|^{1/4} \, \exp \Bigg[ \pm (- 1)^L \, i \int \limits_{t_{\widetilde{A}}}^{t} p_3(x) dx  \Bigg]  \, \varphi_L^{(\mp)}(\textbf{k}, t)   ,
\end{eqnarray}
($L = 1, 2$). Insertion of (\ref{Gl18}) into (\ref{Gl16}) yields for $m = 0$

\begin{equation}
\label{Gl19}
\partial_t \phi^{(j, \mp)}(\textbf{k}, t) - \Omega^{(\mp)}(\textbf{k}, t) \phi^{(j, \mp)}(\textbf{k}, t) = 0   ,
\end{equation}
where the two linearly independent spinor solutions for each chirality state are labeled by j, and where

\begin{equation}
\label{Gl20}
\Omega^{(\mp)}(\textbf{k}, t)  = \left( \begin{array}{rr} 0 \  \  \  \  \  \  \  \    &  \mathcal{P}^{(\mp)}(\textbf{k}, t)  \\   - [\mathcal{P}^{(\mp)}(\textbf{k}, t)]^{\ast}  &  0 \  \  \  \  \  \   \end{array} \right) 
\end{equation}
with  ($t_{\widetilde{A}} \geq 0$)

\begin{equation}
\label{Gl21}
\mathcal{P}^{(\mp)}(\textbf{k}, t)  = \pm (ip_1 + p_2)  \exp \Bigg[ \mp 2i \int \limits_{t_{\widetilde{A}}}^{t} p_3(x) dx  \Bigg]   .
\end{equation}
The system (\ref{Gl19}) and (\ref{Gl20}) has the following useful property: Given any solution $\Phi = 
(\Phi_1,\Phi_2)^T$, then $\Psi = (\Phi_2^\ast, - \Phi_1^\ast)^T$ represents a second solution orthogonal to $\Phi$ w.r.t. the hermitean scalar product $(\Phi, \Psi)  \, = \, \sum_{L = 1}^{2} \Phi_L^\ast \Psi_L$.

Solving (\ref{Gl19}) one gets with (\ref{Gl15}), (\ref{Gl18}) the pertaining four bispinor solutions:

\begin{equation}
\label{Gl22}
\psi_\textbf{k}^{(j, \mp)}(\textbf{x},t) = c^{(j, \mp)}_\textbf{k} \, e^{i\textbf{k} \textbf{x}} \left( \begin{array}{c} \varphi^{(j, \mp)}(\textbf{k}, t)  \\ \mp \, \varphi^{(j, \mp)}(\textbf{k}, t)  \end{array} \right)   ,
\end{equation}
$(j  = 1, 2)$. Note that (\ref{Gl21}) implies $\mathcal{P}^{(-)}(\textbf{k}, t) \rightarrow \mathcal{P}^{(+)}(\textbf{k}, t)$ when $\textbf{k} \rightarrow  - \textbf{k}$. Hence, to obtain all solutions of (\ref{Gl19}), one must only find the two solutions of, say, the negative chirality case. The positive chirality solutions are then:

\begin{equation}
\label{Gl23}
\phi^{(j, +)}(\textbf{k}, t) = \phi^{(j, -)}(- \textbf{k}, t).
\end{equation}

An appropriate scalar product for Dirac-spinors is defined by

\begin{equation}
\label{Gl24}
\left\langle u,v \right\rangle = \int\limits_{\Sigma} \ast F_{u,v}
\end{equation}
with $\Sigma$ a spacelike Cauchy hypersurface and $\ast$ the duality operator. Furthermore, $F_{u,v}$ is given by

\begin{equation}
\label{Gl25}
F_{u,v}(x) \equiv [F_{u,v}(x)]_{\mu} \, \Theta^{\mu} := \overline{u(x)} \,  \gamma_{\mu} v(x) \,\Theta^{\mu} ,
\end{equation}
where $u$ and $v$ are solutions of (\ref{Gl14}), and $\overline{u}$ denotes the Dirac adjoint. The integral on the r.h.s. of equation (\ref{Gl24}) is independent of the choice of the hypersurface $\Sigma$, which can be seen as follows: The 4-form
\begin{equation}
\ast \delta F_{u,v} = - \, \Theta^0 \wedge \Theta^1 \wedge \Theta^2 \wedge \Theta^3 \{e_{\mu}([F_{u,v}]^{\mu}) + C^{\mu}_ {\ \mu \nu} [F_{u,v}]^{\nu} \} \nonumber
\end{equation}
(with $\delta$ the codifferential or adjoint operator) vanishes, since

\begin{equation}
\label{Gl26}
e_{\mu}([F_{u,v}]^{\mu}) + C^{\mu}_ {\ \mu \nu} [F_{u,v}]^{\nu} = 0 .
\end{equation}
The desired result follows then as a consequence of Gauss` theorem: $\int_{\partial V} \ast F_{u,v} = \int_V \ast \delta F_{u,v} = 0$. Eq. (\ref{Gl26}) appears as Lorenz gauge condition in Maxwell's theory, if instead of (\ref{Gl25}) the 1-form $A = A_\mu(x) \Theta^\mu$ is used \cite{Wollensak2}. The inner product (\ref{Gl24}) is sesquilinear, satisfies $\left\langle u,v \right\rangle = \left\langle v,u \right\rangle^{\ast}$ and is positive definite. Hence, eq.s ($ \ref{Gl24}  $) and ($ \ref{Gl25}  $) define a hermitean scalar product, which will be used to define the normalization condition for Dirac-spinors:

\begin{equation}
\label{Gl26a}
< \psi_{\textbf{k}}^{(j, c)}, \ \psi_{\textbf{k}^{\prime}}^{(j^{'},c^{'})}>  \, = \, \delta_{j j^{'}} \, \delta_{c c^{'}} \, \delta(\textbf{k} \, - \, \textbf{k}^{\prime})  
\end{equation}
with $j, j^{'} = 1, 2; \, c, c^{'} = -, +$.
The covectors $\Theta^\mu$ are related to the covectors of the coordinate frame via $\Theta^\mu = D^\mu_{\ l}(x)\, dx^l$ (greek indices refer to the orthonormal, latin ones to the coordinate frame), where the $D^\mu_{\ l}(x)$ denote the vierbein fields. Correspondingly, one finds for the vectors $e_\mu$, $\partial_l$ the relation $e_\mu = [D^{- 1} (x)]^l_{\ \mu}\, \partial_l$. In a coordinate frame, eq.(\ref{Gl25}) reads: $F_{u,v}(x) = [\widetilde{F}_{u,v}(x)]_l dx^l \equiv \overline{u(x)} \,  \widetilde{\gamma}_l(x) v(x) \, dx^l$ with $\widetilde{\gamma}_l(x) = D^\mu_{\ l}(x)\, \gamma_\mu$, and ($ \ref{Gl26}  $)
takes then the form $\partial_l[\sqrt{|g|}(\widetilde{F}_{u,v})^l] = 0$. 
\\
\section*{3. Weyl TEO in Planar BI Universes}
For simplicity we specialize now to pBI spacetimes. Since we are also interested in such backgrounds at very early times, we assume the scale factors to be of power law type. This is motivated by the fact that close to the singularity, BI spacetimes can be well approximated by suitable vacuum Kasner geometries \cite{Lif}. Hence, in the following we consider background geometries with scale factors $\alpha_1 \equiv \alpha_2 := t^{\nu}, \ \alpha_3 := t^{1 - \mu} \, (\mu > 0)$, i.e. (\ref{Gl1}) simplifies to

\begin{equation}
\label{Gl27}
ds^2 = dt^2 -  t^{2 \nu} (dx^1)^2 -  t^{2 \nu} (dx^2)^2 -  t^{2 - 2 \mu} (dx^3)^2   .
\end{equation}
Nevertheless, it is extremely difficult to find exact solutions of (\ref{Gl19}) even for those simple anisotropic backgrounds defined by (\ref{Gl27}).\footnote{A rare example for exactly soluble models is the stiff-fluid class treated in sec. 4.3.}

We put now $k_3 \neq 0$, because the case $k_3 = 0$ is exactly soluble. Defining

\begin{equation}
\label{Gl28}
k_{\pm} = (k_2 \pm ik_1) \, e^{ \pm  2ik_3 \, t^{\mu}_{\widetilde{A}}/\mu }   ,
\end{equation}
one gets with (\ref{Gl21})

\begin{equation}
\label{Gl29}
\mathcal{P}^{(-)}(\textbf{k}, t) = k_+ \, t^{- \nu} e^{- 2ik_3 \, t^{\mu}/\mu} .
\end{equation}
In what follows, we seek negative chirality solutions of eq. ($ \ref{Gl19}  $). To this end we define the (negative chirality) operator $\hat{\Omega}_{\textbf{k}}^{(-)}$ acting on the function $\xi$

\begin{equation}
\label{Gl30}
\hat{\Omega}_{\textbf{k}}^{(-)}[\xi] = \xi_A + \int \limits_{t_A}^t  \Omega^{(-)}(\textbf{k}, y) \, \xi(\textbf{k}, y) dy   ,
\end{equation}
with $\Omega^{(-)}$ given in (\ref{Gl20}), $\xi_A = \hat{\Omega}_{\textbf{k}}^{(-)}\xi(\textbf{k}, t_A)$ and $t_A \geq t_{\widetilde{A}} \geq 0$. With the help of this operator, the problem of solving the system ($ \ref{Gl19}  $) can be cast into the equivalent fixed-point problem $\hat{\Omega}^{(-)}_{\textbf{k}}[\phi^{(l, -)}] = \phi^{(l, -)} \,  (l = 1, 2).$ It can be shown that such a fixed point exists, and that it is the only one. Moreover, for any arbitrary continous function $\xi^{(l)}$ holds: $\{ \hat{\Omega}_ {\textbf{k}}^{(-)} \}^n [\xi^{(l)}](t) \stackrel{n \rightarrow \infty}{\rightarrow} \phi^{(l, -)}(\textbf{k}, t)$ \cite{Weissinger}, that is:
\begin{equation}
\label{Gl32}
\begin{aligned}
\phi^{(l, -)}(\textbf{k}, t) \, = \, & \Big [ \,  1_2 +  \sum_{n=1}^{\infty}  \int\limits_{t_A}^t dt_1 \int\limits_{t_A}^{t_1} dt_2 ...\int\limits_{t_A}^{t_{n-1}} dt_n \, \Omega^{(-)}(\textbf{k}, t_1)
\\
&\times \Omega^{(-)}(\textbf{k}, t_2) ... \Omega^{(-)}(\textbf{k}, t_n) \, \Big] \, \xi^{(l)}_A  ,
\end{aligned}
\end{equation}
where $(1_2)_{LM} \equiv \delta_{LM}$. With $\xi^{(l)}_A = \phi^{(l, -)}({\textbf{k}}, t_A)$, eq. (\ref{Gl32}) can be written in terms of a time-ordered exponential:

\begin{equation}
\phi^{(l, -)}(\textbf{k}, t) \, = \, T \exp \left( \int\limits_{t_A}^t \Omega^{(-)}(\textbf{k}, y) dy \right) \phi^{(l, -)}({\textbf{k}}, t_A)    .    \nonumber
\end{equation}
This expression satisfies by definition eq. (\ref{Gl19}). It is reminiscent of a result by Tsamis and Woodard \cite{Tsamis} who determined massless scalar field solutions in fFLRW backgrounds with arbitrary expansion factor.

We introduce now

\begin{equation}
\label{Gl33}
s(t) = t^{\mu}
\end{equation}
($\mu > 0, \, s_A := s(t_A)$), and define the parameter

\begin{equation}
\label{Gl34}
\delta = (1 - \nu)/\mu.
\end{equation}
This parameter can vanish or even take on negative values provided $t_A > 0$. Eq. (\ref{Gl32}) can now be written in compact form:

\begin{equation}
\label{Gl35}
\phi^{(l, -)}(\textbf{k} ,t)  = K_{\textbf{k}}^{(-)}(t|t_A) \, \phi^{(l, -)}(\textbf{k}, t_A)   ,
\end{equation}
where the negative chirality Weyl TEO $K_{\textbf{k}}^{(-)}$ reads

\begin{equation}
\label{Gl36}
K_{\textbf{k}}^{(-)}(t|t_A)  =  \sum_{n=0}^{\infty} \left( \begin{array}{rr}  I_n(s)   & \  0 \ \ \  \\  \  0 \  \  \  &  I^{\ast}_n(s)   \end{array} \right) \, \mathcal{H}^n_{\textbf{k}}   , \ \ \  \    \mathcal{H}_{\textbf{k}} : = 
\left( \begin{array}{rr} \  \  0 \  \  &  \frac{k_+}{\kappa}  \\   - \frac{k_-}{\kappa}  & \  \ 0 \   \end{array} \right)   ,
\end{equation}
with $I_0(s) \equiv 1,  \, \mathcal{H}^2_{\textbf{k}} = - 1_2$, and

\begin{equation}
\label{Gl37}
I_n(s)  = \left(  \frac{\kappa s^\delta}{\mu} \right)^n  \int\limits_{\sigma_A}^1 d\sigma_1 \int\limits_{\sigma_A}^{\sigma_1} d\sigma_2 ...\int\limits_{\sigma_A}^{\sigma_{n-1}}    d\sigma_n 
\prod\limits_{l = 1}^{n} \sigma_l^{\delta -1} 
e^{i (- 1)^l \sigma_l \, \tau}
\end{equation}
($\sigma_A \leq a \leq b \leq 1$), and where

\begin{equation}
\label{Gl411}
\tau(s) \, = \, 2 k_3  s/\mu  \, \equiv \, 2 k_3 t^\mu/\mu  ,
\end{equation}
\begin{equation}
\sigma_A(s) = s_A/s  \equiv  \tau_A/\tau  \geq 0 ,
\end{equation}
\begin{equation}
\label{Gl38}
\kappa = \sqrt{k_1^2 + k_2^2} \equiv \sqrt{k_+ \, k_-} , \  \ \eta_{ \delta} = \frac{\kappa^2}{2 |k_3|^{2 \delta }}   .
\end{equation}
It now turns out that, provided $\mu, \, \delta > 0$, the quantity $\eta_{ \delta}$ plays the role of an expansion parameter at late times $\tau$, i.e. it can be taken to be small when $|\tau|$ becomes large. Observe that $|\tau| \gg 1$ does not imply: $s \gg 1$. It can also be satisfied for any $s  \gtrsim \mu > 0$ by  $|k_3| \gg 1$. As a consequence, $|k_3|$ can be regarded as a large and thus $\eta_\delta$ as a small quantity, if  $|\tau| \gg 1$.

Equations (\ref{Gl35}) - (\ref{Gl37}) represent the two linearly independent \textit{exact} negative chirality spinor solutions of a massless spin-$\frac{1}{2}$ field in spacetimes (\ref{Gl27}). This result is the basis for the recently introduced parameter transformation (PT) \cite{Wollensak1}, which transforms an exact (and approximate, resp.) massless bispinor solution $\psi$ propagating in a background $ds^2$, into a new exact (and approximate, resp.) solution $\psi^{'}$ with different background $ds^{' 2}$. The new line element $ds^{' 2}$, and the new solution $\psi^{'}$ belong to the same equivalence class of line elements as $ds^2$, and to the same equivalence class of exact (and approximate, resp.)  solutions as  $\psi$. The PT is given by:

\begin{equation}
\label{Gl39}
\nu \rightarrow \nu^{'} = 1 - a (1 - \nu), \, \mu \rightarrow \mu^{'} = a \mu  , 
\end{equation}
$(a \neq 0$ real). It can be viewed as a map $(M, \boldsymbol{g}) \, \rightarrow \,(M^{'}, \boldsymbol{g^{'}})$, where  w.r.t. the coordinate frame

\begin{equation}
\label{Gl40}
g_{\alpha \beta}(x) = \mathrm{diag} \, (1, - x_0^{2 \nu}, - x_0^{2 \nu}, - x_0^{2 - 2 \mu})    ,
\end{equation}
and

\begin{equation}
g^{'}_{\alpha \beta}(x)  = \mathrm{diag} \, (1, - x_0^{2 \nu^{'}}, - x_0^{2 \nu^{'}}, - x_0^{2 - 2 \mu^{'}})   .
\end{equation}
$M$, $M^{'}$ are differentiable manifolds endowed with metrics $\boldsymbol{g}$ and $\boldsymbol{g^{'}}$, and dim $M^{'} \equiv$ dim $M = n$ (with $M = M^{'}$ and here: $n = 4$).

As an alternative one could employ the nonlinear diffeomorphism $f^{(a)}: (M, \boldsymbol{g}) \rightarrow (\widetilde{M}, \boldsymbol{\widetilde{g}})$, defined by:

\begin{equation}
\label{Gl40a}
x_0 \mapsto \widetilde{x}_0 = x_0^a, \  \  \ x_j \mapsto \widetilde{x}_j = a x_j   ,
\end{equation} 
$(j = 1, 2, 3; \,  x_0 \equiv t > 0  , \, a \neq 0)$ together with the additional restriction:

\begin{equation}
\label{Gl40aa}
\boldsymbol{\tilde{k} \, \tilde{x}} = \boldsymbol{k \, x}   .
\end{equation} 
$\widetilde{M}$ denotes a third differentiable manifold with metric $\boldsymbol{\widetilde{g}}$, and dim $\widetilde{M} \equiv n$. In practice, it is often more convenient to utilize $f^{(a)}$ instead of (\ref{Gl39}). Applying this diffeomorphism to (\ref{Gl27}) and (\ref{Gl22}), one obtains the new line element $d\widetilde{s}^2$ and spinor $\widetilde{\psi}$, which are Weyl-related to their PT-transformed counterparts, i.e.: $ \, d\widetilde{s}^2 = \Lambda^2 ds^{' 2}$ and $ \widetilde{\psi} = \Lambda^{- 3/2} \psi^{'}$, with $\Lambda(x) := a x_0^{a - 1}$. In more mathematical terms, the relationship between $f^{(a)}$ and PT (\ref{Gl39}) is established through the dual map $f^{(a)^\ast}$, which applied to $\boldsymbol{g^{'}}$ yields $\Lambda^2  \boldsymbol{g}$, that is, $f^{(a)}$ can be viewed as a conformal map (s. App. C).

The above presented results are exact. Since an exact computation of eq.s (\ref{Gl36}) and (\ref{Gl37}) can only be carried out for a few special cases such as $\delta = 1$ or $k_3 = 0$ (see sec.s 4.1and 4.2), we will focus attention on the determination of an analytical expression for the approximate Weyl TEO. Before starting we note that according to (\ref{Gl37}) holds: $I_n(s) \equiv I^{(1)}_n(\widetilde{\tau})  I^{(2)}_n(\tau)$ with $\widetilde{\tau}(s) := \kappa s^\delta/\mu \propto \sqrt{\eta_\delta} |\tau|$. In  all asymptotic calculations that follow it is always understood that $|\tau| \gg 1 \, (s \gg \mu/2 |k_3|, k_3 \neq 0)$ applies, so that the phase factors in (\ref{Gl37}) oscillate rapidly. Clearly, $|\tau| \gg 1$ does not necessarily imply $|\widetilde{\tau}| \gg 1$, and vice versa. Furthermore, in all that follows we choose $s_A$ so, that it is certainly within the small-time regime. Hence, $s_A \lesssim  \mu/2 |k_3|$. We define now $\epsilon_l := 1 - \sigma_l$, so that $0 \leq \epsilon_1 \leq \epsilon_2 ...\leq \epsilon_n \leq 1$, and write for $l \neq n$: 

\begin{equation}
\label{Gl40ab}
f_1(\sigma_l) := \sigma_l^{\delta - 1} \, \approx \, e^{(1 - \delta) (1 - \sigma_l)} =: f_2(\sigma_l)   ,
\end{equation}
where for $\delta < 1$ holds: $f_R(\sigma_{l + 1}) \geq f_R(\sigma_l), \, (R = 1, 2)$, that is we put

\begin{equation}
\label{Gl41}
\prod\limits_{l = 1}^{n} \sigma_l^{\delta -1} \approx \sigma_n^{\delta - 1} e^{(1 - \delta) (n - 1)} \exp \left[(\delta - 1) \sum\limits_{l = 1}^{n - 1} \sigma_l \right]    .
\end{equation}
Insertion of (\ref{Gl41}) into (\ref{Gl37}) gives then:

\begin{equation}
\label{Gl42}
\widetilde{I}_n(s) = e^{\delta - 1} \left(  \frac{\kappa s^\delta}{e^{\delta - 1} \mu} \right)^n   \int\limits_{\sigma_A}^1 d\sigma_1  ...\int\limits_{\sigma_A}^{\sigma_{n-1}} d\sigma_n \sigma_n^{\delta -1} e^{i (- 1)^n  \sigma_n \tau}
\prod\limits_{l = 1}^{n-1}  e^{(\delta -1) \sigma_l + i (- 1)^l \sigma_l \tau}
\end{equation}
with $\widetilde{I}_0(s) \equiv 1$. Substituting $\widetilde{I}_n(s)$ for $I_n(s)$ in (\ref{Gl36}) leads to the approximate negative chirality Weyl TEO. Note that in (\ref{Gl41}) we have retained the most divergent factor $\sigma_n^{\delta - 1}$ ($\delta < 1$). It helps to make eq.s (\ref{Gl36}) and (\ref{Gl42}) a sensible approximation for the exact TEO at early and late times. This can be easily verified for small times, because $I_n(s) \equiv \widetilde{I}_n(s)$ for $n = 0, 1$. These are already the most dominant terms when $\Delta s = s - s_A \ll 1$. Moreover, in this case holds: $\sigma_A \approx 1$, so that $\sigma_l \approx 1$ and hence $\epsilon_l \ll 1$ for all $l$. It follows for $n \geq 2$: $\widetilde{I}_n(s) \to I_n(s)$ if $\Delta s \to 0$. The  asymptotic case is dicussed in App. A.

To perform the computation of (\ref{Gl42}) and summation in (\ref{Gl36}) we use

\begin{equation}
\label{GlA2}
\widetilde{I}_n(s) = s^{\delta n}  e^{- i\frac{\tau(s)}{2}}
\int\limits_{0}^{1 - \sigma_A}  dz \, h_n(1 - \sigma_A - z; s)
\left( g_{0, n - 1}  \ast g_{1, n - 1}  \ast ... \ast g_{n - 1, n - 1}  \right) (z;s)
\end{equation}
with

\begin{equation}
\label{GlA3}
\begin{aligned}
g_{l, j}(z;s) & = \exp \left\lbrace \left[ (1 - \delta)(j - l) + i (- 1)^{l} \, \tau(s)/2   \right] z \right\rbrace  ,
\\
h_j(z;s)  & = (z + \sigma_A)^{\delta - 1} \exp \left\lbrace   i (- 1)^{j} \, \tau(s)  \, (z + \sigma_A)/2 \right\rbrace ,
\end{aligned}
\end{equation}
where  $(g \ast h) (z;s) := \int^z_0 g(z - \zeta; s) h(\zeta; s) d\zeta$. Applying a Laplace transformation and then its inverse, the convolution in (\ref{GlA2}) assumes the form:

\begin{equation}
\label{GlA4}
\left( g_{0, n - 1}  \ast g_{1, n - 1}  \ast ... \ast g_{n - 1, n - 1}  \right) (z;s) \,
= \, \sum\limits_{l = 1}^{n} e^{{\cal{G}}_{l, n}(s) \, z} {\prod\limits_{m = 1, \, m \neq l}^{n}} \, \frac{1}{\Delta {\cal{G}}_{l, m}(s)}
\end{equation}
with $\Delta {\cal{G}}_{l, m} := {\cal{G}}_{l, j} - {\cal{G}}_{m, j}$ and ${\cal{G}}_{l, j}(s): = - i (- 1)^{l} \, \tau(s)/2 + (1 - \delta)(j - l)$. Insertion of (\ref{GlA2}), (\ref{GlA4}) into (\ref{Gl36}) yields then for the approximate TEO a matrix expression, where in the entries terms of the form

\begin{equation}
\sum\limits_n \left( \frac{i \kappa s^{\delta}}{\mu} \right)^n \sum\limits_{l = 1}^{n} e^{{\cal{G}}_{l, n}(s) \, z} {\prod\limits_{m = 1, \, m \neq l}^{n}} \, \frac{1}{\Delta {\cal{G}}_{l, m}(s)}   \nonumber 
\end{equation}
appear. Exploiting properties of ${\cal{G}}_{l, j}$ and $\Delta {\cal{G}}_{l, m}$ such as ${\cal{G}}_{k, j} = {\cal{G}}_{k, l} + (1 - \delta)(j - l)$ and $\Delta {\cal{G}}_{l + 2k, m + 2k} = \Delta {\cal{G}}_{l, m}$ ($l < m$), and rearranging the summands, we find for the diagonal and off-diagonal entries of $K_{\textbf{k}}^{(-)}$ after some algebra

\begin{equation}
\label{GlA5}
\begin{aligned}
\sum\limits_{n = 1}^{\infty} \left( \frac{i \kappa s^{\delta}}{\mu} \right)^{2n} \sum\limits_{l = 1}^{2n} \frac{e^{{\cal{G}}_{l, 2n}(s) \, z}}{{\prod\limits_{m = 1, \, m \neq l}^{2n}}  \, \Delta {\cal{G}}_{l, m}(s)} \, =& \, - \, e^{{\cal{G}}_{1, 1}(s) \; z} \; S_1 \;  S_4^{\ast}\ +\  e^{{\cal{G}}_{2, 2}(s) \; z} \; S_2 \; S_3^{\ast}  
\\
\sum\limits_{n = 1}^{\infty} \left( \frac{i \kappa s^{\delta}}{\mu} \right)^{2 n - 1} \sum\limits_{l = 1}^{2 n - 1} \frac{e^{{\cal{G}}_{l, 2n - 1}(s) \, z}}{{\prod\limits_{m = 1, \, m \neq l}^{2n - 1}} \, \Delta {\cal{G}}_{l, m}(s)} \, 
&= \, 
e^{{\cal{G}}_{1, 1}(s) \; z} \; S_1 \; S_3 \ +\  e^{{\cal{G}}_{2, 2}(s) \; z} \; S_2 \; S_4  
\end{aligned}
\end{equation}
where

\begin{equation}
\begin{aligned}
S_1 &= \Gamma(\lambda) \left( x/2 \right)^{\lambda^{\ast}} J_{- \lambda^{\ast}}(x)  ,  \  \   &S_3 = \Gamma(\lambda^{\ast}) \left( X/2 \right)^{\lambda} J_{- \lambda}(X)  ,
\\
S_2 &= - i \Gamma(\lambda^{\ast}) \left( x/2 \right)^{\lambda} J_{\lambda^{\ast}}(x)  , \  &S_4 = i \Gamma(\lambda) \left(X/2 \right)^{\lambda^{\ast}} J_{\lambda}(X)  ,
\end{aligned}   \nonumber
\end{equation}
with $J_{\pm \lambda}$ denoting Bessel`s function, $\Gamma$ the Gamma function \cite{Abramowitz}, and

\begin{equation}
\label{Gl52}
x(s) = \frac{\kappa s^{\delta}}{\mu (1 - \delta)}  ,
\ \ \
X(s; z) =  e^{(1 - \delta) z} x(s)  ,
\ \ \
\lambda(s) = \frac{1}{2} \, + \, \frac{i\,\tau(s)}{2 (1 - \delta)}   .
\end{equation}
We use (\ref{Gl36}), (\ref{GlA2}), (\ref{GlA4}), (\ref{GlA5}) and eventually arrive at the approximate Weyl TEO in background spacetimes (\ref{Gl27})

\begin{equation}
\label{Gl53}
K_{\textbf{k}}^{(-)}(t| t_A) \, = \,
\left( \begin{array}{rr} \left( K_{\textbf{k}}^{(-)}(t|t_A) \right)_{11}     &  \left( K_{\textbf{k}}^{(-)}(t|t_A) \right)_{12}  \\ 
\\  - \left( K_{\textbf{k}}^{(-)}(t|t_A) \right)^{\ast}_{12}  & \left( K_{\textbf{k}}^{(-)}(t|t_A) \right)_{11}^{\ast}    \end{array} \right)
\end{equation}
with

\begin{equation}
\label{Gl54}
\begin{aligned}
\left( K_{\textbf{k}}^{(-)}(t|t_A) \right)_{11} & = 1 - \frac{\kappa s^\delta}{\mu} \, \int\limits_0^{1 - \sigma_A} dz \, \left( \frac{e^{\lambda^\ast z}}{1 - z} \right)^{1 - \delta} \, \frac{R(z; x)}{Z(0; x)}  \ ,
\\
\left( K_{\textbf{k}}^{(-)}(t|t_A) \right)_{12} & =  \frac{k_+ s^\delta}{\mu} \, e^{- i \tau} \int\limits_0^{1 - \sigma_A} dz  \, \left( \frac{e^{\lambda z}}{1 - z} \right)^{1 - \delta} \, \frac{Z(z; x)}{Z(0; x)}  ,
\end{aligned}
\end{equation}
and

\begin{equation}
\label{Gl55}
\begin{aligned}
R(z; x) &= J_{- \lambda}(\mathcal{D} x) J_{\lambda} (\mathcal{D} X) - J_{\lambda}(\mathcal{D} x) J_{- \lambda}(\mathcal{D} X)  ,
\\
Z(z; x) &=  J_{- \lambda}(\mathcal{D} x) J_{- \lambda^{\ast}} (\mathcal{D} X) + J_{\lambda}(\mathcal{D} x) J_{\lambda^{\ast}}(\mathcal{D} X)  .
\end{aligned}
\end{equation}
Here, $\delta \leq 1$, and  for the time being we set $\mathcal{D} \equiv  1$. Note that

\begin{equation}
\label{Gl55a}
x \, \Gamma(\lambda^{\ast}) \, \Gamma(\lambda)/2  \, = \, \pi x/2 \cosh[\pi \tau/2 (1 - \delta)]  \equiv  1/Z(0; x)   .
\end{equation}
We emphasize that in the above calculation no further approximation was used, that is eq.s (\ref{Gl53}) - (\ref{Gl55}) represent for $\mathcal{D} = 1$ the exact result of (\ref{Gl36}), (\ref{Gl42}). But there is the following drawback of approximation (\ref{Gl42}): When $s_A = 0 \ (\sigma_A = 0)$, it does for $n \geq 2$ not correctly reproduce the divergent behavior, if $\delta$  approaches zero. This problem can be fixed by introducing the normalization constant $\mathcal{D}_1 := \int_0^1 d\sigma f_1(\sigma) \, / \, \int_0^1 d\sigma f_2(\sigma)$. The combination with a further fine-tuning constant $\mathcal{D}_2 \approx 1 \, (0< \delta \leq 1)$ gives then

\begin{equation}
\label{Gl60}
\mathcal{D}(\delta)  =  \mathcal{D}_1(\delta)  \mathcal{D}_2(\delta) \ \stackrel{\delta \rightarrow 1}{\rightarrow} \ 1    ,
\end{equation}
where $\mathcal{D}_1 = (\delta^{- 1} - 1) / (e^{1 - \delta} - 1)$ and $\mathcal{D}_2 := 2$ $_1F_1(1; 1 + \delta; 1 - \delta)  / (e^{1 - \delta} + 1)$. We can undo this approximation by setting again $\mathcal{D} = 1$ in eq.s (\ref{Gl55}).

The asymptotic expansions of eq.s (\ref{Gl54}) are derived in App. B, and in App. C it is shown that the approximate solutions (\ref{Gl53}) - (\ref{Gl55}) tend (with $\mathcal{D} \equiv 1$) in the PT limiting case to the correct fFLRW result, just as the exact solutions do.
\\
\section*{4. Exact and Approximate Results}
We study first three exactly solvable models which serve as testing ground for eq.s (\ref{Gl35}), (\ref{Gl53}) - (\ref{Gl55}):  massless fermions moving in hypersurfaces $x^3 =$ const. ($\delta$ arbitrary), in a fFLRW ($\delta = 1$) and in an anisotropic stiff-fluid background ($\delta = 1/2$). The fourth example with planar anisotropic Kasner background ($\delta = 1/4$) can only be solved approximately.
\\
\subsection*{4.1. Propagation in hypersurfaces}
The limiting case $k_3 \to 0$ yields spinors propagating in timelike hypersurfaces $x^3 = \mathrm{const.}$ With (\ref{Gl19}), (\ref{Gl20}) and (\ref{Gl29}) follows:
\begin{equation}
\partial_t \phi_1^{(j, -)}= k_+ \, t^{- \nu} \, \phi_2^{(j, -)}  ,\ \ \
\partial_t \phi_2^{(j, -)}= - k_- \, t^{- \nu} \, \phi_1^{(j, -)}   .    \nonumber
\end{equation}
The exact solutions are given by

\begin{equation}
\label{Gl62}
\phi^{(1, -)}(\textbf{k},t) = A_1 \left( \begin{array}{c} \cos y(t)  \\ - \frac{\kappa}{k_+} \sin y(t) \end{array} \right)   ,
\ \
\phi^{(2, -)}(\textbf{k},t) =  A_2 \left( \begin{array}{c} \sin y(t)  \\ \frac{\kappa}{k_+} \cos y(t) \end{array} \right)
\end{equation}
with $\textbf{k} \equiv (k_1, k_2,0)^T$, and $y = \kappa \, (t^{1 - \nu}  -  t_A^{1 - \nu})/(1 - \nu)$, which can be either directly calculated \cite{Kamke} or by virtue of (\ref{Gl35}) - (\ref{Gl37}) with $I_n(s) \equiv [\kappa (s^{\delta} - s_A^{\delta})/\mu \delta]^n/n!$. These solutions oscillate with time-dependent frequency $\omega(t) = \kappa t^{- 1} ln(t/t_A)$, if $\nu = 1$, and $\omega(t) = \kappa t^{ - \nu}/(1 - \nu)$ otherwise. 

Approximation (\ref{Gl41}) cannot be expected to make sense if $k_3 = 0$, but ignoring this for the moment we obtain with $\mathcal{D} = 1$:

\begin{equation}
\frac{R(z; x)}{Z(0; x)} \, \stackrel{k_3 \rightarrow 0}{\rightarrow}  \, \frac{\sin[x (e^{(1 - \delta)z} - 1)]}{e^{(1 - \delta) z/2} }  ,\ \  \  \
\frac{Z(z; x)}{Z(0; x)} \, \stackrel{k_3 \rightarrow 0}{\rightarrow}  \, \frac{\cos[x (e^{(1 - \delta)z} - 1)]}{e^{(1 - \delta) z/2} }   \nonumber  ,
\end{equation}
and executing the integrals in (\ref{Gl54}) leads with

\begin{equation}
\label{Gl64}
\xi(s) \, = \,  x(s) (e^{(1 - \delta) (1 - \sigma_A(s))} - 1) 
\end{equation}
to:

\begin{equation}
\label{Gl65}
K_{\textbf{k}}^{(-)}(t| t_A)  \approx  \left( \begin{array}{rr} \cos[\xi(s)]     & \frac{k_+}{\kappa}\, \sin[\xi(s)]  \\  - \frac{k_-}{\kappa}\, \sin[\xi(s)]  &  \cos[\xi(s)]   \end{array} \right) .
\end{equation}
The closer $\delta$ or $\sigma_A$ is to unity, the better this result will be. In fact, eq. (\ref{Gl65}) would represent the exact TEO following directly from eq.s (\ref{Gl36}), (\ref{Gl37}), if one replaces $\xi$ with $\xi_{exact} \equiv y  = x (1 - \delta) (1  -  \sigma_A^\delta)/\delta$. Using

\begin{equation}
\phi^{(1, -)}(\textbf{k},t_A) = A_1 \left( \begin{array}{c} 1  \\ 0 \end{array} \right)   ,   \ \ \   
\phi^{(2, -)}(\textbf{k},t_A) =  \frac{\kappa A_2}{k_+}  \left( \begin{array}{c} 0  \\ 1 \end{array} \right)   , \nonumber
\end{equation}
one arrives by virtue of eq.s (\ref{Gl35}) and (\ref{Gl65}) again at (\ref{Gl62}) with, however, $\xi$ substituted for $y$. But for $\delta \rightarrow 0$ ($\nu \rightarrow 1$), eq. (\ref{Gl64}) does not reproduce the correct logarithmic behavior. 

If one applies in addition approximation (\ref{Gl60}), one gets

\begin{equation}
K_{\textbf{k}}^{(-)}(t| t_A)  \approx  \left( \begin{array}{rr} 1 - \frac{1}{\mathcal{D}} + \frac{\cos[\mathcal{D}  \xi]}{\mathcal{D}}     & \frac{k_+}{\kappa}\, \frac{\sin[\mathcal{D}  \xi]}{\mathcal{D}}   \\  - \frac{k_-}{\kappa}\, \frac{\sin[\mathcal{D}  \xi]}{\mathcal{D}}  &  1 - \frac{1}{\mathcal{D}} + \frac{\cos[\mathcal{D}  \xi]}{\mathcal{D}}   \end{array} \right)  , \nonumber
\end{equation}
which is unitary for $\mathcal{D} = 1$ and $\mathcal{D} \rightarrow \infty$, i.e. (\ref{Gl60}) is only sensible if $k_3 \neq 0$.
\\
\subsection*{4.2. Conformally flat FLRW background}
We consider spin-$\frac{1}{2}$ fields in a radiation-dominated universe, that is eq. (\ref{Gl27}) with $\mu =  \nu = 1/2 \ (\delta = 1)$. In this case, eq.s (\ref{Gl54}) are exact and read\footnote{One gets $all$ exact TEOs belonging to the class $\delta = 1$ by use of (\ref{Gl40a}) and (\ref{Gl40aa}), s. App. C!}

\begin{equation}
\label{Gl70}
\begin{aligned}
\left( K_{\textbf{k}}^{(-)}(t|t_A) \right)_{11} &= e^{- 2 i k_3 (\sqrt{t} - \sqrt{t_A}) } \left\lbrace 
\cos[2k (\sqrt{t}  - \sqrt{t_A})] + i  \, \frac{k_3}{k} \,  \sin[2k (\sqrt{t} - \sqrt{t_A})] \right\rbrace
\\
\left( K_{\textbf{k}}^{(-)}(t|t_A) \right)_{12} &= \frac{k_+}{k} \, 
e^{- 2 i k_3 (\sqrt{t} + \sqrt{t_A})}\, \sin [2 k (\sqrt{t} - \sqrt{t_A})]  .
\end{aligned}
\end{equation}
To compare with previous results, we set $t_{\widetilde{A}} = t_A = 0$ and use

\begin{equation}
\phi^{(j, -)}(\textbf{k}, 0) =  \left( \begin{array}{c} 1  \\  \frac{(- 1)^j \, k\, \mathrm{sign} k_3 \, - \, k_3 }{k_1 - i k_2}  \end{array} \right)    \nonumber
\end{equation}
as initial condition. The negative chirality bispinor solutions are given by

\begin{equation}
\label{Gl72}
\begin{aligned}
\psi_{\textbf{k}}^{(j, -)}(\textbf{x},t)& \ = \ c^{(j, -)}_\textbf{k} \, e^{i\textbf{k} \textbf{x}} \left( \begin{array}{c} \varphi^{(j, -)}(\textbf{k}, t)  \\ - \, \varphi^{(j, -)}(\textbf{k}, t)  \end{array} \right)   ,
\\
\varphi^{(j, -)}(\textbf{k}, t)& \, = \, |g(t)|^{-1/4} \,  e^{2 i (- 1)^j \,  k \, \mathrm{sign}k_3 \, \sqrt{t}}
\left( \begin{array}{c} 
1  \\ 
 \frac{(- 1)^j \, k\, \mathrm{ sign} k_3 \, - \, k_3 }{k_1 - i k_2} 
\end{array} \right)   ,
\end{aligned}
\end{equation}
($|g| \equiv t^3$). Analogously, the positive chirality solutions are owing to (\ref{Gl23}):

\begin{equation}
\label{Gl74}
\begin{aligned}
\psi_{\textbf{k}}^{(j, +)}(\textbf{x},t)& \ = \ c^{(j, +)}_\textbf{k} \, e^{i\textbf{k} \textbf{x}} \left( \begin{array}{c} \varphi^{(j, +)}(\textbf{k}, t)  \\  \, \varphi^{(j, +)}(\textbf{k}, t)  \end{array} \right)   ,
\\
\varphi^{(j, +)}(\textbf{k}, t)& = |g(t)|^{-1/4} \,  e^{- 2 i (- 1)^j \,  k \, \mathrm{sign}k_3 \, \sqrt{t}}
\left( \begin{array}{c}  1  \\  \frac{(- 1)^j \, k\, \mathrm{ sign} k_3 \, - \, k_3 }{k_1 - i k_2} 
\end{array} \right) .
\end{aligned}
\end{equation}
Normalizing with (\ref{Gl26a}) leads to:
\begin{equation}
\label{Gl78}
c^{(1, -)}_\textbf{k} \, = \, \frac{\sin \zeta }{\sqrt{2} \, (2 \pi)^{3/2}}  , \ \ \ \  c^{(2, -)}_\textbf{k} \, = \, c^{(1, -)}_\textbf{k}  \ \cot \zeta ,\ \ \ \   c^{(j, +)}_\textbf{k} \, = \, c^{(j, -)}_\textbf{k} ,  \nonumber
\end{equation}
where $\sin \zeta :=   \sqrt{ (k - |k_3|)/2k}, \    \cos \zeta :=   \sqrt{ (k + |k_3|)/2k}$. The four solutions (\ref{Gl72}) and (\ref{Gl74}) are orthogonal w.r.t. ($ \ref{Gl24}  $). Suitable linear combinations of these chirality eigenspinors yield the outcome of ref. \cite{Barut}, for example:

\begin{equation}
\sin \zeta \ \, \psi_{\textbf{k}} ^{(1, +)}  (\textbf{x}, t)   \ + \
 \cos \zeta \ \, \psi_{\textbf{k}}^{(2, -)}(\textbf{x}, t) \ = 
\frac{t^{- 3/4} \, e^{i\textbf{k} \textbf{x} \, + \, 2 i  k \, \mathrm{sign}k_3 \, \sqrt{t}}}{\sqrt{2} (2 \pi)^{3/2}} \, 
\left( \begin{array}{c} 
\mathrm{sign} k_3
\\
0
\\
- \frac{k_3}{k}
\\
- \frac{k_1 + i k_2}{k}
\end{array} \right)  .  \nonumber
\end{equation}
\\
\subsection*{4.3. Anisotropic stiff-fluid model}
Next we study exact solutions of Einstein`s field equations when the material content is described by a "perfect fluid" \cite{Kramer}. A special case is given by line element ($ \ref{Gl27}  $) with $\mu = 1,\, \nu = 1/2 \,\ (\delta = 1/2$). Using (\ref{Gl19}), (\ref{Gl20}) and (\ref{Gl29}) one obtains for the negative chirality case ($j = 1, 2$):

\begin{equation}
\label{Gl81}
\partial_t \phi_1^{(j, -)} = \frac{k_+}{\sqrt{t}} \, e^{- i \tau}  \,  \phi_2^{(j, -)}  , \ \
\partial_t \phi_2^{(j, -)} = - \frac{k_-}{\sqrt{t}} \, e^{i \tau}  \,  \phi_1^{(j, -)}   ,
\end{equation}
where $\tau \equiv 2 k_3 t$. Exact solutions of this system are given by \cite{Kamke}, \cite{Pimentel}:\footnote{By means of (\ref{Gl40a}) and (\ref{Gl40aa}), one obtains the exact solutions of $all$ models belonging to the class $\delta = 1/2$ \cite{Wollensak1}!}

\begin{equation}
\label{Gl82}
\begin{aligned}
\phi_1^{(1, -)}(\textbf{k}, t) \, = \,& (- i \tau)^{- \frac{1}{4}} \, e^{\frac{- i \tau}{2}} \, W_{- \frac{1}{4} - i \eta; \frac{1}{4}}( i \tau)   ,
\\
\phi_2^{(1, -)}(\textbf{k}, t) \, = \,& \frac{\sqrt{- 2i k_3}}{k_+} \, (- i \tau)^{- \frac{3}{4}} \, e^{\frac{i \tau}{2}} \,  
[i \eta \, W_{- \frac{1}{4} - i \eta; \frac{1}{4}}(i \tau) \ - \  W_{ \frac{3}{4} - i \eta; \frac{1}{4}}(i \tau) ]   ,
\end{aligned}
\end{equation}
and

\begin{equation}
\label{Gl83}
\begin{aligned}
\phi_1^{(2, -)}(\textbf{k}, t) \, = \,& (- i \tau)^{- \frac{1}{4}} \, e^{\frac{- i \tau}{2}} \, W_{ \frac{1}{4} + i \eta; \frac{1}{4}}(- i \tau)   ,
\\
\phi_2^{(2, -)}(\textbf{k}, t) \, = \,& i \, \frac{k_- \, \mathrm{sign}k_3}{\sqrt{ 2i k_3}} \, (- i \tau)^{- \frac{3}{4}} \, e^{\frac{i \tau}{2}} \, 
\left[ W_{ \frac{1}{4} + i \eta; \frac{1}{4}}(- i \tau)  + q_\eta \, W_{ - \frac{3}{4} +  i \eta; \frac{1}{4}}(- i \tau) \right]   ,
\end{aligned}
\end{equation}
where $W_{\mu; \nu}$ denotes Whittaker`s function, $q_\eta := i \eta - 1/2$, and $\eta := \eta_{\frac{1}{2}} \, \mathrm{sign}k_3  \equiv \kappa^2/2k_3$. The bispinor solutions read owing to (\ref{Gl22}) with (\ref{Gl18}), (\ref{Gl23}) (and $|g(t)| \equiv t^2$)

\begin{equation}
\label{Gl85}
\psi^{(j, \mp)}(\textbf{x},t) \, = \,  d^{(j, \mp)}_\textbf{k} \, e^{i\textbf{k} \textbf{x}} \, |g(t)|^{- 1/4}
\left( \begin{array}{c}  e^{\frac{i \tau}{2}}  \, \phi_1^{(j, \mp)}(\textbf{k}, t)
\\
 e^{\frac{- i \tau}{2}}  \, \phi_2^{(j, \mp)}(\textbf{k}, t)
\\
- e^{\frac{i \tau}{2}}  \, \phi_1^{(j, \mp)}(\textbf{k}, t)
\\
- e^{\frac{- i \tau}{2}}  \, \phi_2^{(j, \mp)}(\textbf{k}, t)
\end{array} \right)  ,
\end{equation}
($j = 1, 2$). One obtains from the exact solutions (\ref{Gl82}), (\ref{Gl83})  for $|\tau| \ll 1$

\begin{equation}
\label{Gl86}
\begin{aligned}
\phi^{(1, -)}(\textbf{k},t) \, = \, & \sqrt{\pi} \, e^{i \frac{\pi}{4} \mathrm{sign} k_3} \, 
\left( \begin{array}{c}  \frac{1}{\Gamma(1 + i \eta)} \, - \, \frac{2 \sqrt{i \tau}}{\Gamma(\frac{1}{2} + i \eta)} + O(\tau)
\\
- \frac{\sqrt{2i k_3}}{k_2 + i k_1} \left[ \frac{1}{\Gamma(\frac{1}{2} + i \eta)} \ - \ \frac{2 \sqrt{i \tau}}{\Gamma(i \eta)} + O(\tau) \right]
\end{array} \right)   ,
\\
\phi^{(2, -)}(\textbf{k},t) \, = \, & \left( \begin{array}{c}  \frac{\Gamma(\frac{1}{2})  }{\Gamma(\frac{1}{2} - i \eta)} \, - \, \frac{2 \sqrt{- i \tau}}{\Gamma (- i \eta) } + O(\tau)
\\
i \, \frac{(k_2 - i k_1) \, \mathrm{sign} k_3 }{\sqrt{2i k_3}} \frac{\Gamma(\frac{1}{2})}{\Gamma(1 - i \eta)} \left[1 \ - \ \frac{ 2 \,  \Gamma(1 - i \eta) \, \sqrt{- i \tau}}{\Gamma(\frac{1}{2} - i \eta)} + O(\tau) \right]
\end{array} \right)   ,
\end{aligned}
\end{equation}
and for $|\tau| \gg 1$

\begin{equation}
\label{Gl87}
\begin{aligned}
\phi^{(1, -)}(\textbf{k},t) \, \sim \, &  \, e^{i \frac{\pi}{4} \mathrm{sign} k_3} \, (i \tau)^{- i\eta} 
\left( \begin{array}{c}  \ \frac{e^{- i \tau}}{\sqrt{i \tau}} \, \left[ 1 + \frac{1 + 3i \eta - 2 \eta^2}{ -2 i \tau} + O(\frac{1}{\tau^2}) \right]
\\
- \frac{\sqrt{2i k_3}}{k_2 + ik_1} \left[ 1 +  \mathrm{sign} k_3 \, \frac{\eta - 2 i \eta^2}{- 2 i \tau} + O(\frac{1}{ \tau^2})  \right]
\end{array} \right)   ,
\\
\phi^{(2, -)}(\textbf{k},t) \, \sim \, & \, (- i \tau)^{ i\eta}
\left( \begin{array}{c}  1 + \frac{i \eta + 2 \eta^2}{- 2 i \tau} + O(\frac{1}{\tau^2})
\\
i \, \frac{(k_2 - ik_1) \, \mathrm{sign} k_3}{\sqrt{2i k_3}} \frac{e^{i \tau}}{\sqrt{- i \tau}} \left[ 1 - \frac{(1 - i \eta)(1 - 2i \eta)}{- 2 i \tau} + O(\frac{1}{\tau^2}) \right]
\end{array} \right) .
\end{aligned}
\end{equation}
We compare this outcome with the TEO result (\ref{Gl54}). For $|\tau| \gg 1$ one has:

\begin{equation}
\label{Gl88}
\begin{aligned}
\left( K_{\textbf{k}}^{(-)}(t|0) \right)_{11} \,  \sim  & \, [1 + O( \eta)]   ,
\\
\left( K_{\textbf{k}}^{(-)}(t|0) \right)_{12} \,  \sim & \ \frac{k_2 + ik_1}{\sqrt{- 2i k_3}} \, \left[
- i \sqrt{\pi} \, \mathrm{sign} k_3 \, + \, \frac{e^{- i \tau}}{\sqrt{- i \tau}} \, [1 + \, O(\eta)] \right] ,
\end{aligned}
\end{equation}
following from (\ref{GlB14}). Choosing as general initial conditions ($j = 1, 2$)

\begin{equation}
\phi^{(j, -)}(\textbf{k},0) \, = \,\left( \begin{array}{c}
\phi_1^{(j, -)}(\textbf{k},0)
\\
\phi_2^{(j, -)}(\textbf{k},0)
\end{array} \right)  \nonumber
\end{equation}
with at least one component satisfying $\phi_L^{(j, -)} (\textbf{k},0) \neq  0$, and utilizing  (\ref{Gl35}) one gets for $|\tau| \rightarrow \infty$ the two asymptotic Weyl spinor solutions. Conversely, one can also deduce the initial condition from the asymptotic expressions (\ref{Gl88}). To see this explicitly, we use the asymptotic conditions:

\begin{equation}
\label{Gl89}
\phi_1^{(1, -)}(\textbf{k},t) \ \stackrel{|\tau| \rightarrow \infty}{\rightarrow} \ 0  , \  \  \   \
\phi_2^{(1, -)}(\textbf{k},t) \ \stackrel{|\tau| \rightarrow \infty}{\rightarrow} \ \mathrm{const.} \neq 0 ,
\end{equation}
which are compatible with the asymptotic result (\ref{Gl87}) of the exact result. As a consequence, the spinor components $\phi_L^{(1, -)}(\textbf{k},t) \ (L = 1, 2)$ must obey a second condition at $t = 0$ following from  (\ref{Gl35}), (\ref{Gl88}), (\ref{Gl89}):

\begin{equation}
\label{Gl90}
\phi_2^{(1, -)}(\textbf{k},0) \, = \, - \, \frac{\sqrt{2i k_3}}{k_2 + ik_1} \, \frac{1}{\Gamma(\frac{1}{2})} \, [1 + O( \eta)]  \, \phi_1^{(1, -)}(\textbf{k},0) .
\end{equation}
This condition represents according to (\ref{Gl86}) to lowest order in $\eta$ the correct initial value at $t = 0$ of the exact solution $\phi^{(1, -)}(\textbf{k},t)$. 

The asymptotic expansion of $\phi^{(1, -)}(\textbf{k}, t)$ reads with  (\ref{Gl35}), (\ref{Gl88}) and (\ref{Gl90})

\begin{equation}
\label{Gl91}
\frac{\phi^{(1, -)}(\textbf{k},t)}{\phi_1^{(1, -)}(\textbf{k},0)}  \, = \, \frac{1}{\sqrt{\pi}}  \,
\left( \begin{array}{c}
\frac{e^{- i \tau}}{\sqrt{i \tau}} \, [1 + O(\eta)]
\\
- \frac{\sqrt{2i k_3}}{k_2 + ik_1} \, \left[ 1 + O(\eta)  + O\left(  \frac{\eta}{\sqrt{- i \tau}}  \right) \right]
\end{array} \right)   .
\end{equation}
The determination of the asymptotic expansion of $\phi^{(2, -)}(\textbf{k},t)$ by use of (\ref{Gl88}) proceeds along the same lines as before: We demand instead of (\ref{Gl89}) as asymptotic condition (guaranteeing orthogonality of $\phi^{(1, -)}$ and $\phi^{(2, -)}$)

\begin{equation}
\label{Gl92}
\phi_1^{(2, -)}(\textbf{k},t) \ \stackrel{|\tau| \rightarrow \infty}{\rightarrow} \ \mathrm{const.} \neq 0 ,  \   \  \  \
\phi_2^{(2, -)}(\textbf{k},t) \ \stackrel{|\tau| \rightarrow \infty}{\rightarrow} \ 0   ,
\end{equation}
which implies analogous to (\ref{Gl90}) the second condition:

\begin{equation}
\label{Gl93}
\phi_2^{(2, -)}(\textbf{k},0) \, = \, \frac{i (k_2 - ik_1) \, \mathrm{sign}k_3}{\sqrt{2i k_3}} \, \Gamma \left( \frac{1}{2} \right)  \, [ 1 + O(\eta) ]  \, \phi_1^{(2, -)}(\textbf{k},0).\
\end{equation}
Again, this result is to lowest order in $\eta$ the correct initial condition for the exact solution $\phi^{(2, -)}(\textbf{k},t)$. For $|\tau| \gg 1$ one obtains

\begin{equation}
\label{Gl94}
\frac{ \phi^{(2, -)}(\textbf{k},t)}{ \phi_1^{(2, -)}(\textbf{k},0)}  \, =  \, \left( \begin{array}{c} 1 + O(\eta) + O\left(\frac{\eta}{\sqrt{i \tau}}  \right)
\\
i \, \frac{(k_2 - ik_1) \mathrm{sign}k_3}{\sqrt{2i k_3}} \,  \frac{e^{i \tau}}{\sqrt{- i \tau}}  \, [ 1 + O(\eta) ]
\end{array} \right) .
\end{equation}
Comparison of ($ \ref{Gl91}  $), ($ \ref{Gl94}  $) with the asymptotic expansions ($ \ref{Gl87}  $) of the exact results shows agreement apart from the prefactors $(\pm i \tau)^{\mp i \eta}$. These can be disregarded in the asymptotic limit (s. also App. A). For $t \rightarrow 0$ holds

\begin{equation}
\label{Gl95}
K_{\textbf{k}}^{(-)}(t| 0) \, = \,
\left( \begin{array}{rr} 1 \ \ \   &  \frac{k_+}{\mu \, \delta} \, s^{\delta}  \\ -  \frac{k_-}{\mu \, \delta} \, s^{\delta}   &  1 \ \ \   \end{array} \right) \, [1 + O(s^{2 \delta})] ,
\end{equation}
following from eq.s (\ref{Gl53}) - (\ref{Gl55}). Insertion of the exact initial values $\phi^{(j, -)}(\textbf{k},0)$ into (\ref{Gl35}) yields the correct behavior at early times, eq.s (\ref{Gl86}).
\\
\subsection*{4.4. Anisotropic axisymmetric Kasner model}
In this model the background is described by line element (\ref{Gl27}) with $\mu = 4/3$, $\nu = 2/3$ ($\delta = 1/4$). This exact vacuum solution evolves e.g. for $t \rightarrow 0$ as the preinflationary limiting case of a pBI background geometry \cite{Gm}.

The system of differential equations reads (s. eq.s (\ref{Gl19}), (\ref{Gl20}) and (\ref{Gl29}), $\tau \equiv 3 k_3 t^{4/3}/2$)

\begin{equation}
\label{Gl96}
\partial_t \phi_1^{(j, -)} = \frac{k_+}{t^{2/3}} \, e^{- i \tau}  \, \phi_2^{(j, -)}  ,\ \ 
\partial_t \phi_2^{(j, -)} = - \frac{k_-}{t^{2/3}} \, e^{i \tau} \, \phi_1^{(j, -)} .
\end{equation}
For simplicity we set: $t_{\widetilde{A}}  =  t_A = 0$ and get from (\ref{Gl35}), (\ref{Gl53}), (\ref{GlB9}) and (\ref{GlB12}):

\begin{equation}
\label{Gl102}
\phi_1^{(1, -)}(\textbf{k},0) \, = \, - \, (3/4)  \, \Gamma(1/4) \, k_+ \, \left( 3 i k_3/2  \right)^{- 1/4}  \, \phi_2^{(1, -)}(\textbf{k},0)   ,
\end{equation}
where we imposed again the asymptotic behavior (\ref{Gl89}) and used $\eta_\delta \ll 1$. In the same way one finds with the asymptotic condition (\ref{Gl92})

\begin{equation}
\label{Gl104}
\phi_1^{(2, -)}(\textbf{k},0) \, = \, \frac{4}{3} \, \frac{1}{ \Gamma(\frac{1}{4})} \, \frac{1}{k_-} \, \left[  \left( \frac{3 i}{2} \, k_3  \right)^\frac{1}{4}  \right]^\ast \, \phi_2^{(2, -)}(\textbf{k},0) .
\end{equation}
To determine for $|\tau| \gg 1$ the spinors $\phi^{(1, -)}$ and $\phi^{(2, -)}$, we substitute eq. (\ref{Gl102}) into (\ref{Gl35}), (\ref{Gl53}), (\ref{GlB9}), (\ref{GlB12}) and get

\begin{equation}
\label{Gl1052}
\begin{aligned}
\phi_1^{(1, -)}(\textbf{k},t)   \, \sim  \, &A^{(1,-)}_1(\textbf{k}) \, \frac{e^{- i \tau}}{|\tau|^{1 - \delta}}  \,  [1 \, + \, O(\eta_\delta  \tau^{2 \delta - 1}) ] \, [ 1 + O(\eta_\delta)] \, \phi_2^{(1, -)}(\textbf{k},0)   ,
\\
\phi_2^{(1, -)}(\textbf{k},t) \, \sim \, &[  1  -  i \, \mathrm{sign}k_3  \left(  \mu/2 \right)^{2 \delta - 2} (\mathcal{D}/2)  \left\lbrace \,  _1F_1(1;1+\delta;1-\delta)/\delta    \, + \, O(\eta_\delta)  \right\rbrace
\\
&\times  \eta_\delta  |\tau|^{2 \delta - 1} \, + \, O(|\tau|^{ \delta - 1}) ]   \, [ 1 \, + \,  O(\eta_\delta) ]  \, \phi_2^{(1, -)}(\textbf{k},0)   ,
\end{aligned}
\end{equation}
with $A^{(1,-)}_1(\textbf{k}) :=  \frac{i}{2} \mathrm{sign}k_3 \, \left( \frac{\mu}{2}  \right)^{\delta - 1} k_+/|k_3|^\delta$, 
where the spinor $\phi_2^{(1, -)}(\textbf{k},0)$ on the r.h.s of eq.s (\ref{Gl1052}) denotes the lowest-order term of the $\eta_\delta$-expansion of the exact solution $\phi_2^{(1, -)}(\textbf{k},t)$ at $t = 0$. In the same manner one obtains

\begin{equation}
\label{Gl1053}
\begin{aligned}
\phi_1^{(2, -)}(\textbf{k},t) \, \sim \, &[  1  +  i \, \mathrm{sign}k_3  \left(  \mu/2 \right)^{2 \delta - 2} \mathcal{D} \left\lbrace \,  _1F_1(1;1+\delta;1-\delta) \, + \, O(\eta_\delta)  \right\rbrace
\\
&\times  \eta_\delta  |\tau|^{2 \delta - 1}/2\delta \, + \, O(|\tau|^{ \delta - 1}) ]   \, [ 1 \, + \,  O(\eta_\delta) ]  \, \phi_1^{(2, -)}(\textbf{k},0)   ,
\\
\phi_2^{(2, -)}(\textbf{k},t)   \, \sim & \,
A^{(2,-)}_2(\textbf{k}) \, \frac{e^{ i \tau}}{|\tau|^{1 - \delta}}  \,  [1 \, + \, O(\eta_\delta  \tau^{2 \delta - 1}) ] \, [ 1 + O(\eta_\delta)] \, \phi_1^{(2, -)}(\textbf{k},0)   ,
\end{aligned}
\end{equation}
with $A^{(2,-)}_2 = - (A^{(1,-)}_1)^\ast$. As above $\phi_1^{(2, -)}(\textbf{k},0)$ represents the lowest-order term of the $\eta_\delta$-expansion of the exact solution $\phi_1^{(2, -)}(\textbf{k},t)$ at $t = 0$. The bispinor solutions are again given by (\ref{Gl85}) (with $|g(t)| \equiv t^2$). 

The late- and early-time solutions of (\ref{Gl96}) have recently been determined by use of a different approach \cite{Wollensak1}. Agreement with the above solutions was found in the small- and large-time regimes. The correct matching of these two regimes, however, is only possible with the help of the TEO results.
\\
\section*{5. Expansion about Conformally Flat Universes}
The class of conformally flat line elements is defined by (\ref{Gl27}) satisfying $\delta = 1$. In the following we will now investigate spacetimes with small deviations off conformal flatness, defined by $\delta :=  1 - \epsilon$ ($\epsilon$  a small positive parameter). This does not imply that we are dealing with weak anisotropy only. Let us take the quantity $\Delta H/H$\footnote{This quantity squared is up to a constant the usual mean anisotropy parameter.} as measure of anisotropy in a general pBI universe ($\alpha_1 \equiv \alpha_2 \neq \alpha_3$ in (\ref{Gl1})), with $H := \sum_j \partial_t \ln \alpha_j/3$ the average Hubble parameter and $\Delta H := \partial_t \ln \alpha_1 - \partial_t \ln \alpha_3$ the difference of the directional Hubble parameters. Then, except for $\mu$ close to unity, it follows for the special case (\ref{Gl27}), that the parameter $\epsilon$ is related to (weak) anisotropy via

\begin{equation}
|\Delta H/H| \approx \mu \epsilon /|1 - \mu|   .
\end{equation}
But if $\mu \approx 1, \, |\Delta H/H|$ will indeed be large and even tend to infinity when $\mu \rightarrow \mu_0 = (1 - 2 \epsilon/3)^{- 1}$, while $\epsilon \ll 1$ is still maintained.
Note also that, in contrast to the case $\delta = 1$, no two line elements (\ref{Gl27}) belonging to the same equivalence class $\delta = 1 - \epsilon$ are conformally equivalent.

To derive suitable expressions for $R$, $Z$ given by (\ref{Gl55}), we utilize the product formula for Bessel functions \cite{Erdelyi}
 
\begin{equation}
\label{Gl106}
J_{\mp \lambda}(x) \, J_{\pm \lambda}(x e^{\epsilon z}) \,  = \,  e^{\pm \lambda \epsilon z} \, \sum\limits_{n = 0}^{\infty} \, \frac{[\frac{x}{2} (1 - e^{2 \epsilon z})]^n}{n!} \, J_{\pm \lambda + n}(x) \, J_{\mp \lambda}(x)
\end{equation}
with  $2 \lambda \equiv 1 + i \tau/\epsilon$ and $x \equiv \kappa s^{1 - \epsilon}/\mu \epsilon$. Owing to (\ref{GlB31}) holds:

\begin{equation}
\label{Gl106a}
\begin{aligned}
J_{\pm \lambda + n}(x) \, J_{\mp \lambda}(x) \, = \,  &\frac{1}{\pi} \int\limits_{- \frac{\pi}{2}}^{\frac{\pi}{2}} dv \, e^{\pm i 2 \lambda v} \, e^{i n v} J_n(2x \, \cos v)   ,
\\
J_{\pm \lambda^\ast + n}(x) \, J_{\pm \lambda}(x)  \, = \,  &\frac{1}{\pi} \int\limits_{- \frac{\pi}{2}}^{\frac{\pi}{2}} dv \, e^{\mp \tau  v/\epsilon} \, e^{ - i n v} J_{n \pm 1}(2x \, \cos v)  .
\end{aligned}
\end{equation}
Splitting the domain of integration and shifting $v \to \frac{\tau}{\epsilon} \, (\frac{\pi}{2} - v)$, we get

\begin{equation}
\label{Gl107}
\begin{aligned}
R(z;s) \, =  \, & \frac{i  \epsilon }{\pi \tau} \, e^{\frac{\pi |\tau|}{2 \epsilon}}   \,
 \sum\limits_{n = 0}^{\infty} \, \frac{[i \, \mathrm{sign} k_3 \, \mathcal{D} x  \frac{1 - e^{2 \epsilon z} }{2} ]^n}{n!} \, 
\\
& \times  \Bigg[ \ (- 1)^{n + 1} \, e^{\lambda \epsilon z}  \int\limits_0^\infty dv \, e^{- v} \, e^{i (n + 1) \epsilon v/\tau} \,J_n[2 \mathcal{D} x \, \sin(\epsilon v/|\tau|)]
\\
& + \, e^{- \lambda \epsilon z}  \int\limits_0^\infty dv \, e^{- v} \, e^{- i (n - 1) \epsilon v/\tau} \,J_n[2 \mathcal{D} x \, \sin(\epsilon v/|\tau|)] \  \Bigg]  ,
\end{aligned}
\end{equation}
where we used (\ref{Gl55}) and (\ref{Gl106}) and neglected terms $\propto e^{- O(|\tau|/\epsilon)} $. Provided $|\tau| \gtrsim 1$, the sines in (\ref{Gl107}) can be linearized, and the integrals are replaced by \cite{Ryzhik}

\begin{equation}
\label{Gl1070}
w_n^{\pm} \, = \, \int\limits_0^\infty dv \, e^{- \rho^\pm_n \, v} \, J_n[\mathcal{D} A_\textbf{k} v]  \,  =  \, \frac{\left( \frac{\mathcal{D} A_\textbf{k}}{\sqrt{(\rho^\pm_n)^2 +  \mathcal{D}^2 A_\textbf{k}^2} \, + \, \rho^\pm_n} \right)^n}{\sqrt{(\rho^\pm_n)^2 + \mathcal{D}^2 A_\textbf{k}^2}}    ,
\end{equation}
with
\begin{equation}
\label{Gl1071}
A_\textbf{k}(s ) \, = \frac{2 x \epsilon}{|\tau |} \, \equiv \, \frac{\kappa}{|k_3|} s^{-  \epsilon}  ,  \ \ \ \rho^\pm_n(\tau) \, = \, 1 - i (1 \pm n) \, \frac{\epsilon}{\tau}  . 
\end{equation}
Insertion of this result into (\ref{Gl107}) gives

\begin{equation}
\label{Gl108}
R(z;s) \, \approx \,   \frac{- \, i \, \frac{ \epsilon}{\pi \tau} \, e^{\frac{\pi |\tau|}{2 \epsilon}}
}{\sqrt{1 + \mathcal{D}^2   A^2_\textbf{k}(s)}}  \,  \left(  e^{\lambda  \epsilon  z \, + \, i \mathcal{Y}_\textbf{k}} \, - \,  e^{-  \lambda  \epsilon  z \, - \, i \mathcal{Y}_\textbf{k}}  \right)
\end{equation}
with
\begin{equation}
\label{Gl109}
\mathcal{Y}_\textbf{k}(z;s) \, = \,\frac{e^{2 \epsilon z}  -  1}{2 \epsilon}  \, y_\textbf{k}(s)  ,  \   \    \
y_\textbf{k}(s) \, =  \,  \frac{\tau(s)}{2} \, \frac{\mathcal{D}^2   \, A^2_\textbf{k}(s)  }{\sqrt{1 + \mathcal{D}^2   A^2_\textbf{k}(s)} \, +  1}  .
\end{equation}
Here, we utilized $\rho^\pm_n \approx 1$, which is permissible for large $|\tau|$: substituting (\ref{Gl1070}) for the integrals in (\ref{Gl107}) we get a series with summands $(\mathfrak{v}_\pm \mathcal{D} A_\textbf{k})^n w_n^{\pm}/n!$ where $\mathfrak{v}_\pm  := \pm i (e^{2 \epsilon z} - 1) \tau/4 \epsilon$. We split now the summation into $\sum_{n = 0}^{N - 1} ... + \sum_{n = N}^{\infty} ...$, with $N(\tau; \epsilon) := \lfloor \mathfrak{g} |\tau|/\epsilon \rfloor \gg 1$ the greatest integer function, and $\mathfrak{g}$ a small positive constant. Then we can replace $w_n^{\pm}$ with $w_n := w_n^{\pm}|_{\rho_n^{\pm} \equiv 1}$ in the first sum, and adding up the first sum to infinity while simultaneously subtracting exactly these additional terms in the second sum gives the desired result plus an additional series with $n \in [N; \infty)$. But this second series is absolutely convergent, hence bounded, and can therefore be made arbitrarily small for sufficiently large $|\tau|$, since $N(\tau; \epsilon) \propto |\tau|$.

Eq.s (\ref{Gl108}), (\ref{Gl55}) put together give

\begin{equation}
\label{Gl110}
\frac{R(z; s)}{Z(0; s)} \, = \, \frac{- i \mathcal{D} A_\textbf{k}(s) \, e^{- \epsilon \lambda^\ast z}}{2 \sqrt{1 \, + \, \mathcal{D}^2  A^2_\textbf{k}(s)}}   \,  [ e^{  \epsilon  z \, + \, i \mathcal{Y}_\textbf{k}} \, - \,  e^{- i  \tau  z \, - \, i \mathcal{Y}_\textbf{k}} ] 
 [1 + O(e^{- \frac{\pi |\tau|}{\epsilon}}) ]   .
\end{equation}
Introducing now

\begin{equation}
\mathcal{Q}_\pm (\textbf{k}; s) \, = \, \frac{\sqrt{1 + \mathcal{D}^2  A^2_\textbf{k}(s) }  \pm 1}{ \sqrt{1 + \mathcal{D}^2  A^2_\textbf{k}(s)}}
\end{equation}
yields with (\ref{Gl106}), (\ref{Gl55}) in the same manner as before

\begin{equation}
\label{Gl111}
Z(z;s) \, = \  \frac{e^{\frac{\pi |\tau|}{2 \epsilon}} }{2 \mathcal{D} \pi x(s)}  \  \Big[  \mathcal{Q}_+ 
  \, e^{- \lambda^\ast \epsilon z \, + \, i \mathcal{Y}_\textbf{k}} \,  + \, \mathcal{Q}_-    \, e^{ \lambda^\ast \epsilon z \, - \, i \mathcal{Y}_\textbf{k}}  \Big] ,
\end{equation}
and finally with (\ref{Gl54})

\begin{equation}
\label{Gl113}
\begin{aligned}
\left( K_{\textbf{k}}^{(-)}(t|t_A) \right)_{11}& \,  = \, 1 \, +  \, i \frac{ y_\textbf{k}(s)}{2 \mathcal{D}} \,  \mathcal{Q}_+  \, [ \,  \mathcal{I}_1(\mathbf{k};s) \,  - \,  \mathcal{I}_2(\mathbf{k};s) \,]  ,
\\
\left( K_{\textbf{k}}^{(-)}(t|t_A) \right)_{12}& \,  = \, \frac{k_+}{2 \mu} \, s^{1 - \epsilon} \, e^{- i \tau} 
\, [ \, \mathcal{Q}_- \, \mathcal{I}^\ast_1(\mathbf{k};s) \, + \,   \mathcal{Q}_+ \mathcal{I}^\ast_2(\mathbf{k};s) \, ]   ,
\end{aligned}
\end{equation}
where

\begin{equation}
\label{Gl115}
\mathcal{I}_1(\mathbf{k};s) \, =  \int\limits_0^{1 - \sigma_A} dz \, 
\frac{e^{\epsilon z \, + \, i \mathcal{Y}_\textbf{k}(z; s)}}{(1 - z)^{\epsilon}}   ,
\ \ \ \
\mathcal{I}_2(\mathbf{k};s) \, =  \int\limits_0^{1 - \sigma_A} dz \, 
\frac{e^{- i \tau z \,-  \, i \mathcal{Y}_\textbf{k}(z; s)}}{(1 - z)^{\epsilon}}  ,
\end{equation}
with $0 \leq \sigma_A(s) \equiv s_A/s \equiv \tau_A/\tau \leq 1$. For $\sigma_A > 0$, one obtains

\begin{equation}
\label{Gl116}
\mathcal{I}_1 \,  =  \, \frac{1}{2 \epsilon} \, \int\limits_0^{\mathcal{E}} dv \, e^{i y_\textbf{k}(s) \frac{v}{2 \epsilon}} \, \Bigg[  1\,  - \,  \epsilon \, \left\lbrace  \frac{\ln (1 + v)}{2 \epsilon} \, + \, \ln \left( 1 - \frac{\ln (1 + v)}{2 \epsilon} \right)  \right\rbrace
 + \, O(\epsilon^2) \Bigg]   ,
\end{equation}
with definition

\begin{equation}
\label{Gl117}
\mathcal{E}(s) \, = \, e^{2 \epsilon [1 - \sigma_A(s)]} - 1 .
\end{equation}
Because $v \leq \mathcal{E} = O( \epsilon) \ll 1$, one has $\ln(1 + v) \approx v$, where convergence of the r.h.s. of (\ref{Gl116}) requires $v/2\epsilon \leq \mathcal{E}/2\epsilon < 1$, and gets eventually \cite{Ryzhik}

\begin{equation}
\label{Gl1171}
\begin{aligned}
\mathcal{I}_1 & (\mathbf{k};s) \,  = \, \frac{e^{i [1 - \mathcal{S}(s)] y_\textbf{k}(s)} - 1}{i y_\textbf{k}(s)}  \, + \, \epsilon \, \frac{e^{i y_\textbf{k}}}{i y_\textbf{k}} \, \Big[ \,  - \left\lbrace 1 - \mathcal{S} \, +  \, \ln(\mathcal{S}) \right\rbrace e^{- i \mathcal{S} y_\textbf{k}} \, + \, \ln(\mathcal{S}) 
\\
&+ \frac{e^{- i \mathcal{S} y_\textbf{k}} - e^{- i y_\textbf{k}}}{i y_\textbf{k}} \,  + \, \widetilde{\mathrm{Ci}}(\mathcal{S} y_\textbf{k})  - i  \, \mathrm{Si}(\mathcal{S} y_\textbf{k})  -  \left\lbrace  \widetilde{\mathrm{Ci}}(y_\textbf{k})  - i  \,  \mathrm{Si}(y_\textbf{k})  \right\rbrace  \Big]   +  O(\epsilon^2)
\end{aligned}
\end{equation}
with constraint $\mathcal{S} > 0$, where

\begin{equation}
\mathcal{S}(s) \, =  \, 1 \ - \, \mathcal{E}(s)/2 \epsilon   ,  \  \  \  \
\widetilde{\mathrm{Ci}}(x) \, =   \ \mathrm{Ci}(x) \, - \,  \log x  .
\end{equation}
Ci(x), Si(x) are the Cosine and Sine integrals, resp. Note that the r.h.s. of (\ref{Gl1171}) is only defined for $\mathcal{S} > 0$! But $\mathcal{S} \to - \, \epsilon$ when $s \to \infty$, and to compute the asymptotic behavior of $\mathcal{I}_1$ we need its analytic continuation:

\begin{equation}
\label{Gl1173}
\begin{aligned}
\mathcal{I}^{(c)}_1 & (\mathbf{k};z) \,  = \, \frac{e^{i [1 - \mathcal{S}(z)] y_\textbf{k}(z)} - 1}{i y_\textbf{k}(z)}  \, + \, \epsilon \, \frac{e^{i y_\textbf{k}}}{i y_\textbf{k}} \, \Big[ \,  - e^{- i \mathcal{S} y_\textbf{k}} \left\lbrace 1 - \mathcal{S}  +   \log (\mathcal{S}) \right\rbrace  +  \log(\mathcal{S}) 
\\
&+   \frac{e^{- i \mathcal{S} y_\textbf{k}} - e^{- i y_\textbf{k}}}{i y_\textbf{k}}   +  \widetilde{\mathrm{Ci}}(\mathcal{S} y_\textbf{k})  - i \,  \mathrm{Si}(\mathcal{S} y_\textbf{k})  -  \left\lbrace  \widetilde{\mathrm{Ci}}(y_\textbf{k})  -  i \,  \mathrm{Si}(y_\textbf{k})  \right\rbrace  \Big]   +  O(\epsilon^2)
\end{aligned}
\end{equation}
where $z$ is now a complex variable, and $\ln[\mathcal{S}(s)]$ has to be replaced with the principal branch of the complex logarithm, $\log [\mathcal{S}(z)]$, with branch cut\footnote{$\mathcal{S}(z)$ is no entire function, in contrast to Si(z) and $\widetilde{\mathrm{Ci}}(z)$!}

\begin{equation}
\left\lbrace  \mathcal{S}(z) = r e^{i \varphi} | r \geq 0, \, \varphi = \alpha - \pi \, (0 < \alpha \ll 1) \right\rbrace . \nonumber
\end{equation}
More precisely, $\mathcal{I}^{(c)}_1(\textbf{k};z)$ represents the analytic continuation of $\mathcal{I}_1(\textbf{k};s)$ along a contour which starts on the real axis at the point $z = s_A > 0$ ($\mathcal{S}(s_A) = 1$), encircles the branch-point $z_o = \frac{s_A}{\epsilon} [1 + O(\epsilon)]$ ($\mathcal{S}(z_o) = 0$) counter-clockwise along a semi-circle with radius $\rho \ll \epsilon$, and after returning back to the real axis at $z_0 + \rho$, proceeds to $+ \, \infty $ (corresponding to $0 > \mathcal{S}(z) \geq - \, \epsilon + O(\epsilon^2)$).

The asymptotic series of $\mathcal{I}^{(c)}_1$ valid for $s \geq \frac{s_A}{\epsilon} + \rho \, (\Rightarrow \mathcal{S} < 0)$ reads \cite{Abramowitz}

\begin{equation}
\label{Gl1174}
\begin{aligned}
\mathcal{I}^{(c)}_1(\mathbf{k};s) \, \sim  \ & \frac{e^{i [1 - \mathcal{S}(s)] y_\textbf{k}(s)} - 1}{i y_\textbf{k}(s)}  \, + \, \epsilon \, \frac{e^{i y_\textbf{k}}}{i y_\textbf{k}} \, \Bigg[ \, - e^{- i \mathcal{S} y_\textbf{k}} \, (1 - \mathcal{S} + \log \mathcal{S}) 
\\
&+ \, \frac{e^{- i \mathcal{S} y_\textbf{k}} - e^{- i y_\textbf{k}}}{i y_\textbf{k}} \, + \, \frac{i e^{- i \mathcal{S} y_\textbf{k}}}{\mathcal{S} y_\textbf{k}} \,  \sum_{n = 0}^{\infty}  (- i \mathcal{S} y_\textbf{k})^{- n}  n!
\\
& - \, \frac{i e^{- i y_\textbf{k}}}{ y_\textbf{k}} \,  \sum_{n = 0}^{\infty}  (- i  y_\textbf{k})^{- n} n! \, \Bigg]   \, + \, O(\epsilon^2)  ,
\end{aligned}
\end{equation}
where $y_\textbf{k}(s) \sim s^{1 - 2 \epsilon}$ when $s \rightarrow \infty$. The second integral $\mathcal{I}_2$ is given by

\begin{equation}
\label{Gl118}
\mathcal{I}_2(\mathbf{k};s) \, = \, \frac{e^{- i \mathcal{A}(\mathbf{k};s)}}{1 - \epsilon} \, \left\lbrace F_{1 - \epsilon}[i \mathcal{A}(\mathbf{k};s)]\, - \, \left( \frac{\tau_A}{\tau} \right)^{1 - \epsilon} \, F_{1 - \epsilon} \left( i \frac{\tau_A}{\tau} \mathcal{A}(\mathbf{k};s)\right)  \right\rbrace
\end{equation}
with $F_{\alpha}(x) := \, _1F_1(\alpha; \alpha + 1; x)$ and

\begin{equation}
\label{Gl119}
\mathcal{A}(\mathbf{k};s)  \, =  \, \tau(s) + y_\textbf{k}(s) \, [1 + O(\epsilon)]  . 
\end{equation}
Insertion of (\ref{Gl1174}) and the asymptotic expansion of (\ref{Gl118}) into (\ref{Gl113}) yields for fixed $\epsilon > 0$ the asymptotic behavior of the Weyl TEO in weakly perturbed conformally flat spacetimes

\begin{equation}
\label{Gl120}
\begin{aligned}
\left( K_{\textbf{k}}^{(-)}(t|t_A) \right)_{11} \,  \sim & \ e^{i [1 - \mathcal{S}(s)] y_\textbf{k}(s)} \, \Big[ 1 - \frac{\epsilon}{2} \Big( 3 -  e^{- i [1 - \mathcal{S}(s)] y_\textbf{k}(s)}  +  O(s_A/s)   +
\\
& \, + 2 \log(-\epsilon + s_A/s  + ...) \Big) + O(\epsilon^2) + O(\tau^{- \epsilon}) \Big]
\end{aligned}
\end{equation}
($\mathcal{D} = 1 + \epsilon/2 + O(\epsilon^2)$), and a similar calculation gives

\begin{equation}
\label{Gl1201}
\left( K_{\textbf{k}}^{(-)}(t|t_A) \right)_{12} \,  \sim  \, - i \, \frac{k_+}{2 k_3} \,e^{i \, \frac{\tau(s)}{4}  A^2_\textbf{k}(s) } \,  \left[ 1 + i (1 - e^{\tau_A}) \right] \, \left[ 1  + O(\epsilon)  + O(\tau^{- \epsilon})  \right]   .
\end{equation}

Some time ago Chimento and Mollerach attempted to obtain massive spinor solutions in BI backgrounds  \cite{Chimento}. They found only two independent solutions, and they also argued that a smooth transition from spinor solutions in a BI background to solutions in a fFLRW background is not possible. In a subsequent work by Castagnino et al. \cite{Castagnino} it was pointed out that the ansatz used in \cite{Chimento} is not the most general one, and provided one can use a separation ansatz for the spinor field one always gets four independent solutions. Furthermore, at least in the case of massless spinor solutions in anisotropic pBI backgrounds, it can explicitly be shown using the above outcome, that for vanishing anisotropy the correct solutions in fFLRW backgrounds are obtained without encountering any discontinuity. Since for fixed $s > 0$ holds
\begin{equation}
A_{\textbf{k}}(s) \stackrel{\epsilon \rightarrow 0}{\rightarrow} \  \frac{\kappa}{|k_3|}, \ \ \
y_\textbf{k}(s)  \stackrel{\epsilon \rightarrow 0}{\rightarrow} \  \frac{\mathrm{sign k_3}}{\mu}  \, (k - |k_3|) s , \  \  \
\mathcal{A}(\textbf{k};s)  \stackrel{\epsilon \rightarrow 0}{\rightarrow} \  \frac{\mathrm{sign k_3}}{\mu} \, (k + |k_3|) s   ,  \nonumber
\end{equation}
and

\begin{equation}
F_{1 - \epsilon}  [i \mathcal{A}(\textbf{k};s)]   \stackrel{\epsilon \rightarrow 0}{\rightarrow} \  \frac{e^{i \frac{\mathrm{sign k_3}}{\mu} \, (k + |k_3|) s} - 1}{i \, \frac{\mathrm{sign k_3}}{\mu} \, (k + |k_3|) s}   ,   \nonumber
\end{equation}
eq.s (\ref{Gl113}) with (\ref{Gl1173}) and (\ref{Gl118}) (these are no asymptotic expressions!) reduce to the matrix elements of the exact Weyl TEO with fFLRW backgrounds:

\begin{equation}
\label{Gl1202}
\begin{aligned}
\left( K_{\textbf{k}}^{(-)}(t|t_A) \right)_{11}  &  \stackrel{\epsilon \rightarrow 0}{\rightarrow} \
e^{- i \frac{k_3}{\mu} \Delta s} \, [ \cos(k \Delta s/\mu) \, + \,i \, (k_3/k) \sin(k \Delta s/\mu) ]   ,
\\
\left( K_{\textbf{k}}^{(-)}(t|t_A) \right)_{12}  &  \stackrel{\epsilon \rightarrow 0}{\rightarrow} \ (k_+/k) \,
e^{- i \frac{k_3}{\mu} (\Delta s + 2 s_A)} \,  \sin(k \Delta s/\mu)    ,
\end{aligned}
\end{equation}
($\Delta s \equiv s - s_A, \, k_+$ as in (\ref{Gl28})). For $\mu = 1/2$ one recovers (\ref{Gl70}). We stress that once the asymptotic expansion of the TEO has been used, $\epsilon$ must stay finite!
\\
\section*{6. Approximate Weyl and Dirac TEO}
Some general properties of the approximate TEO for Weyl spinors and also of the corresponding expression for massless Dirac spinors will now be studied in more detail. We begin with (\ref{Gl53}) and (\ref{Gl54}) and get for small $\Delta t = t - t_A$
\begin{equation}
\label{Gl122}
K_{\textbf{k}}^{(-)}(t| t_A) \, = \,
\left( \begin{array}{rr} 1 \ \ \ \ \  &  \frac{k_+ \, e^{- i \tau}}{t^\nu} \, \Delta t  \\ \frac{- k_- \, e^{ i \tau} }{t^\nu} \, \Delta t   & 1  \ \ \ \ \ \end{array} \right) \, + \, O[(\Delta t)^2] .
\end{equation}
For $t_A = 0$ this Weyl TEO is given by equation (\ref{Gl95}). Owing to (\ref{Gl20}), (\ref{Gl21}), eq. (\ref{Gl122}) can be rewritten as

\begin{equation}
\label{Gl123}
K_{\textbf{k}}^{(-)}(t| t_A)  = 1_2 + \Omega^{(-)} (\textbf{k},t) \, \Delta t  +  O[(\Delta t)^2]   .
\end{equation}
Thus, (\ref{Gl35}) together with (\ref{Gl122}) represent the integrated system (\ref{Gl19}) up to order $\Delta t$. Eq. (\ref{Gl123}) infinitesimally implies:

\begin{equation}
\label{Gl124}
K_{\textbf{k}}^{(-)}(t_C| t_B) \,  K_{\textbf{k}}^{(-)}(t_B| t_A) \, = \, K_{\textbf{k}}^{(-)}(t_C| t_A)  .
\end{equation}
We turn next to the TEO for bispinors $\psi^{(j, -)}$. From (\ref{Gl18}) follows $\varphi^{(j, -)} =  |g|^{- 1/4} \, Q^{- 1} \ \phi^{(j, -)}$, where $Q  =  \mathrm{diag} (Q_{11}, Q_{11}^{\ast})$ with $Q^2_{11}  :=  \mathcal{P}^{(-)} / (i p_1 + p_2),$ and $\mathcal{P}^{(-)}$ defined in (\ref{Gl21}) ($t_A  \geq t_{\widetilde{A}}$). Hence, one obtains

\begin{equation}
\label{Gl126}
K_{\textbf{k}}^{' (-)}(t|t_A) \, = \, |g(t_A)/g(t)|^{1/4} \,  Q^{- 1}(k_3,t) \, K_{\textbf{k}}^{(-)}(t|t_A) \, Q(k_3,t_A)   ,
\end{equation}
which satisfies the analog to (\ref{Gl35}): $\varphi^{(j, -)}(\textbf{k},t) = K_{\textbf{k}}^{' (-)}(t|t_A) \, \varphi^{(j, -)}(\textbf{k},t_A)$. The Dirac TEO for  the negative chirality bispinor (\ref{Gl22}) is given by

\begin{equation}
\label{Gl128}
{\cal{K}}_{\textbf{k}}^{(-)}(t|t_A)  \, = \, 1_2 \, \otimes \, K_{\textbf{k}}^{' (-)}(t|t_A)   ,
\end{equation}
with $1_n$ the $n \times n$ unit matrix. With (\ref{Gl23}) and (\ref{Gl35}) one gets $K_{\textbf{k}}^{(+)} = K_{\textbf{-k}}^{(-)}$, hence

\begin{equation}
\label{Gl129}
K_{\textbf{k}}^{' (+)}(t|t_A) \, = \, |g(t_A)/g(t)|^{1/4} \, Q(k_3,t) \, K_{\textbf{k}}^{(+)}(t|t_A) \, Q^{- 1}(k_3,t_A).
\end{equation}
Thus, as in eq. (\ref{Gl128}) the positive chirality Dirac TEO assumes the form: ${\cal{K}}_{\textbf{k}}^{(+)}  = 1_2 \, \otimes \, K_{\textbf{k}}^{' (+)}$, and by virtue of $K_{\textbf{k}}^{' (+)}  = K_{\textbf{-k}}^{' (-)}$ one gets: ${\cal{K}}_{\textbf{k}}^{(+)} =   {\cal{K}}_{\textbf{-k}}^{(-)}$. The operators ${\cal{K}}_{\textbf{k}}^{(\pm)}$ fulfill a relation analogous to (\ref{Gl124}), and also

\begin{equation}
\label{Gl132}
\Big( {\cal{K}}_{\textbf{k}}^{(\pm)}(t|t_A) \Big)^{\dag}  {\cal{K}}_{\textbf{k}}^{(\pm)} (t|t_A)  =  
\left| \frac{g(t_A)}{g(t)} \right|^{\frac{1}{2}}   \left[ \left| \left( K_{\textbf{k}}^{(-)}(t|t_A) \right)_{11} \right|^2  +  \left|\left( K_{\textbf{k}}^{(-)}(t|t_A) \right)_{12}\right|^2 \right]  1_4    \nonumber
\end{equation}
where at large times follows with (\ref{GlB12})

\begin{equation}
\label{Gl133}
\left| \left( K_{\textbf{k}}^{(-)}(t|t_A) \right)_{11} \right|^2  \, = \, 1  + O(\mathcal{D}  \eta_\delta  |\tau|^{\delta - 1})
\end{equation}
and, utilizing (\ref{GlB9})

\begin{equation}
\label{Gl134}
\left| \left( K_{\textbf{k}}^{(-)}(t|t_A) \right)_{12} \right|^2  \, = \, \Gamma^2(\delta) \, \left( \mu/2 \right)^{2 \delta - 2} \, \eta_\delta
\left[ 1/2 \, + \, O(\eta_\delta |\tau|^{2 \delta - 1})   \right] 
\end{equation}
($0 < \delta < 1/2$ ). As a consequence one obtains with (\ref{Gl24})

\begin{equation}
\label{Gl135}
\begin{aligned}
\Big<   \psi_{\textbf{k}}^{(j, \pm)}&(t), \psi_{\textbf{k}}^{(l, \pm)}(t)  \Big>  \, \equiv \, \left\langle {\cal{K}}_{\textbf{k}}^{(\pm)}(t|t_A) \, \psi_{\textbf{k}}^{(j, \pm)}(t_A) \, , \,  {\cal{K}}_{\textbf{k}}^{(\pm)}(t|t_A) \, \psi_{\textbf{k}}^{(l, \pm)}(t_A)  \right\rangle
\\
&= \,  \int\limits_{\Sigma_{t_A}}  \Theta^1\wedge\Theta^2\wedge\Theta^3 \, \left(  \psi_{\textbf{k}}^{(j, \pm)} \right)^{\dag}  \psi_{\textbf{k}}^{(l, \pm)}  [1 + O(\eta_\delta)  + O(\eta_\delta  |\tau|^{2 \delta - 1}) ]   ,
\end{aligned}
\end{equation}
where

\begin{equation}
\begin{aligned}
\ast F_{{\cal{K}}_{\textbf{k}}^{(\pm)} \, \psi_{\textbf{k}}^{(j, \pm)}, \, {\cal{K}}_{\textbf{k}}^{(\pm)} \, \psi_{\textbf{k}}^{(l, \pm)}} \, = \, 
&\frac{1}{3!} \, \left(  \psi_{\textbf{k}}^{(j, \pm)}(\textbf{x}, t_A) \right)^{\dag} \, 
\left( {\cal{K}}_{\textbf{k}}^{(\pm)}(t|t_A)  \right)^{\dag} \, \gamma^0 
\\
&\times \gamma^{\nu} \, {\cal{K}}_{\textbf{k}}^{(\pm)}(t|t_A) \,  \psi_{\textbf{k}}^{(l, \pm)}(\textbf{x}, t_A) \, \epsilon_{\alpha \beta \gamma \nu} \, \Theta^{\alpha} \wedge \Theta^{\beta} \wedge \Theta^{\gamma}    .   \nonumber
\end{aligned}
\end{equation}
The hypersurface was chosen so that $\Theta^0 = 0$ within $\Sigma_{t_A}$, and we used

\begin{equation}
\epsilon_{\alpha \beta \gamma 0} \, \Theta^{\alpha}\wedge\Theta^{\beta}\wedge\Theta^{\gamma} (t) \, = \, 3! \, \sqrt{|g(t)/g(t_A)|} \, \Theta^1\wedge\Theta^2\wedge\Theta^3 (t_A) .\nonumber
\end{equation}
Clearly, the r.h.s. of (\ref{Gl135}) is just $<\psi_{\textbf{k}}^{(j, \pm)}(t_A), \psi_{\textbf{k}}^{(l, \pm)}(t_A)>  [1 + O(\eta_\delta)  + O(\eta_\delta  |\tau|^{2 \delta - 1})]$. This result is also valid for $\delta = 1/2$. If $\delta = 1 - \epsilon,  \ \epsilon \ll 1$, then one gets with eq.s (\ref{Gl120}), (\ref{Gl1201}):

\begin{equation}
\label{Gl1371}
\frac{\left\langle \psi_{\textbf{k}}^{(j, \pm)}(t) \, , \, \psi_{\textbf{k}}^{(l, \pm)}(t)  \right\rangle}{\left\langle \psi_{\textbf{k}}^{(j, \pm)}(t_A) \, , \, \psi_{\textbf{k}}^{(l, \pm)}(t_A)  \right\rangle}   \, = \, 
1 + O(\eta_1) \,  + O( \epsilon\, \ln \epsilon) + \frac{O(\eta_{1 - \epsilon})  +  O(\epsilon)}{\tau^{\epsilon}}   .  \nonumber
\end{equation}
At early times, one finds

\begin{equation}
\label{Gl139}
\left| \left( K_{\textbf{k}}^{(-)}(t|t_A) \right)_{11}\right|^2 \, + \, \left| \left( K_{\textbf{k}}^{(-)}(t|t_A) \right)_{12}\right|^2 \, = \, 1 + O([\Delta \mathcal{T}]^2) .   \nonumber
\end{equation}
where either holds with (\ref{Gl95}): $\Delta \mathcal{T} = t^{\mu \delta }$ ($t_A \equiv t_{\widetilde{A}} = 0$), or with (\ref{Gl122}): $\Delta \mathcal{T}  \equiv \Delta t = t - t_A$ ($t_A \geq t_{\widetilde{A}} > 0$). Thus,

\begin{equation}
\left\langle \psi_{\textbf{k}}^{(j, \pm)}(t) \, , \, \psi_{\textbf{k}}^{(l, \pm)}(t)  \right\rangle \Big/
\left\langle \psi_{\textbf{k}}^{(j, \pm)}(t_A) \, , \, \psi_{\textbf{k}}^{(l, \pm)}(t_A)  \right\rangle  \, = \, 1  +  O([\Delta \mathcal{T}]^2)  .   \nonumber
\end{equation}
As a result, the approximate Dirac TEOs ${\cal{K}}_{\textbf{k}}^{(\pm)}$ are, apart from small terms of order $\epsilon \ln\epsilon$ and $\eta_\delta$, unitary at early and large times.
\\
\section*{7. Conclusion}
Starting with the formulation of Dirac's equation in anisotropic Bianchi-type-I (BI) spacetimes w.r.t. anholonomic orthonormal frames, and specializing to the massless case in planar BI background spacetimes, an exact expression for the Weyl time-evolution-operator (TEO) of this problem was determined. Based on this solution, a special parameter transformation (PT) can be derived, which is capable of generating exact massless TEO solutions.

The TEO approach also allows the computation of approximate solutions, which behave at early and late times as the exact solutions. The main task is therefore to derive an analytical outcome of the approximate TEO, which can indeed be done for a wide range of the parameter $\delta$. This quantity defines equivalence classes of planar BI background geometries and of the associated TEO solutions. If one knows an approximate TEO solution for a specific $\delta$, then the PT provides all approximate TEOs of the related class.

Comparing the approximate TEO result with solutions of exactly soluble models, we found that it is exact for all conformally flat FLRW models ($\delta = 1$). It also reproduces the short- and large-time behavior of the exact anisotropic stiff-fluid spinor solutions ($\delta = 1/2$).

Especially when no exact solutions are available, the benefit of the TEO technique becomes evident, for example when treating the spin-$\frac{1}{2}$ field in the anisotropic axisymmetric Kasner background ($\delta = 1/4$). 
It could also be shown, that for all models with planar BI backgrounds near conformal flatness ($\delta = 1 - \epsilon$), the smooth transition to conformal flatness is possible.
\\
\section*{Acknowledgments}
The author profited from helpful comments by K.H.Lotze and A.Wipf.
\\
\begin{appendix}
\renewcommand{\theequation}{A\arabic{equation}}
\setcounter{equation}{0}
\section*{Appendix A. Comparison of $I_n$ and $\widetilde{I}_n$ at large times} 
The approximation that leads to eq.s (\ref{Gl54}) and (\ref{Gl55}) is obtained by replacing in eq. (\ref{Gl37}) for $l < n$ $f_1(\sigma_l)$ with $f_2(\sigma_l)$ (s. eq. (\ref{Gl40ab})), that is we are concerned with $\widetilde{I}_n$ given by (\ref{Gl42}) instead of the exact $I_n$. It follows immediately for the lowest order diagonal and off-diagonal entries of the TEO
\begin{equation}
\label{GlA07}
I_0(s) \equiv \widetilde{I}_0(s) \, =  \,1,\  \,   I_1(s) \equiv \widetilde{I}_1(s) \, = \, [\kappa s^\delta/\mu \delta]  \left[ F_\delta(-i \tau) - \sigma_A^\delta \, F_\delta(- i \tau_A) \right]   , \nonumber
\end{equation}
with $F_\delta(x) :=\, _1F_1(\delta; \delta + 1; x) \equiv e^x \, _1F_1(1; \delta + 1; - x)$. The next order diagonal entry in the approximate case is \cite{Ryzhik}

\begin{equation}
\label{GlA08}
\widetilde{I_2}(s) \, = \,  \left(\frac{\kappa s^\delta}{\mu \delta}  \right)^2 \, \frac{\delta}{T} \left[ \frac{F_\delta(\delta - 1) - \sigma_A^\delta  F_\delta([\delta - 1] \sigma_A)}{e^{\delta - 1}} \, - \, \frac{F_\delta(i \tau) - \sigma_A^\delta F_\delta(i \tau_A)}{e^{i \tau}}   \right]     \nonumber
\end{equation}
($T := i \tau + 1 - \delta$), while the exact expression reads

\begin{equation}
\label{GlA09}
\begin{aligned}
I_2(s) \, = \,  \left(\frac{\kappa s^\delta}{\mu \delta}  \right)^2 \, & \Big[\frac{1}{2} \, \mathcal{F}_\delta(- i \tau)  \, -  \, \frac{\sigma_A^{2 \delta}}{2} \, \mathcal{F}_\delta(- i \tau_A) \, - \, \sigma_A^{\delta} \, F_\delta(- i \tau) F_\delta(i \tau_A) 
\\
& + \, \sigma_A^{2 \delta} \, |F_\delta(i \tau_A)|^2  \Big]   ,
\end{aligned}
\end{equation}

with $\mathcal{F}_\delta(x) := \, _2F_2(1, 2 \delta; 2 \delta + 1, \delta + 1; x)$. To derive (\ref{GlA09}) we used: $\delta \, y^{- 2 \delta} \int_{0}^{y}  dx  \, x^{2 \delta - 1}  e^{- i \tau x} \, F_\delta(i \tau x)  =  \mathcal{F}_\delta(- i \tau y)$ \cite{Slater}. Note that $\widetilde{I_2} \to I_2$ for $\delta \to 1$! For simplicity we put now $s_A = 0$ (i.e. $\sigma_A = 0$) and get for $0 < \delta < 1$ the asymptotic expansions \cite{Erdelyi}, \cite{Kim}

\begin{equation}
\label{GlA010}
\begin{aligned}
I_1(s) \,  \equiv \, &\widetilde{I_1}(s)   \sim \mathcal{R}_\textbf{k} (\delta) \, \left\lbrace e^{- i \pi \delta \, \mathrm{sign} k_3} \Gamma(\delta) + \frac{e^{- i \tau}  \, \left\lbrace 1  +  O([- i \tau]^{- 1}) \right\rbrace}{(- i \tau)^{1 - \delta}}   \right\rbrace   ,
\\
I_2(s) \sim \, &\mathcal{R}_\textbf{k}^2 (\delta) \, \left\lbrace
 \frac{1}{1 - 2 \delta}  \left[  C_\delta  + \frac{1}{(- i \tau)^{1 - 2 \delta}} \right] +  \Gamma(\delta) \, \frac{e^{- i \tau} \left\lbrace 1  +  O([- i \tau]^{\delta - 1}) \right\rbrace }{(- i \tau)^{1 - \delta}}  \right\rbrace   ,
\\
\widetilde{I_2}(s) \sim \, &\mathcal{R}_\textbf{k}^2 (\delta) \, \left\lbrace - \frac{_1F_1(1; 1 + \delta; 1 - \delta)/\delta}{ (- i \tau)^{1 - 2 \delta} }  \,  + \,  \Gamma(\delta) \, \frac{e^{- i \tau }  \left\lbrace 1  +  O([- i \tau]^{\delta - 1}) \right\rbrace  }{(- i \tau)^{1 - \delta}}  \right\rbrace ,
\end{aligned}   
\end{equation}
with $C_\delta = [\Gamma(\delta)  e^{- i \pi \delta \, \mathrm{sign} k_3} ]^2   \, (\frac{1}{2} - \delta)/\cos(\pi \delta)$, and

\begin{equation}
\label{GlA123}
\mathcal{R}_\textbf{k} := \kappa (i \mu/2 k_3)^\delta/\mu
\end{equation}
a constant of order $\sqrt{\eta_\delta}$. The time-dependent terms of $I_2, \, \widetilde{I_2}$ agree, apart from the constant factors in front of the terms $\propto \tau^{2 \delta - 1}$. A constant term, however, exists only in the expression for $I_2$. It adds an $O(\eta_\delta)$-term to $I_0$. 

The $\tau^{2 \delta - 1}$-terms in (\ref{GlA010}) are problematic since they diverge for $\delta \geq 1/2 \ (I_2)$ and $\delta > 1/2 \ (\widetilde{I_2})$, if $|\tau| \to \infty$. They cannot be canceled by higher order terms since $I_n = O(\mathcal{R}_\textbf{k}^n) \equiv O(\eta_\delta^{n/2}) = \widetilde{I_n}$. This could cause problems e.g. with the unitarity of the Weyl TEO. These can be avoided by demanding that powers of these terms must show up in every $I_n$ and $\widetilde{I_n}$ in such a way that all these terms add up to a phase factor $e^{i \omega(\tau[t]) \, t}$ with $\eta_\delta$-dependent frequency $\omega$. This requirement is also motivated by the approximate TEO results of this work, see below.
In this context it is instructive to study the special case $\delta = 1/2 \, (\mu = 1)$ a little more closely. We find with $\tau \equiv 2 k_3 t$
\begin{equation}
\label{GlA011}
I_2(s) \sim \, - i \eta \left[\log(i \tau) - \psi(1/2) - \Gamma(1/2) \, \frac{e^{- i \tau}}{\sqrt{- i \tau}} + O(1/i \tau) \right]  ,
\end{equation}
where $\mathcal{R}_\textbf{k}^2 = i \eta \equiv i \eta_{\frac{1}{2}} \, \mathrm{sign}k_3  \equiv i \kappa^2/2k_3$, and

\begin{equation}
\label{GlA012}
\widetilde{I_2}(s) \sim \, - i \eta \left\lbrace 2 \, _1F_1(1; 3/2; 1/2) - \Gamma(1/2) \, \frac{e^{- i \tau}}{\sqrt{- i \tau}} + O(1/i \tau) \right\rbrace   .
\end{equation}
Note that $2 \, _1F_1(1; 3/2; 1/2) \approx - \psi(1/2)$. The essential difference to (\ref{GlA011}) consists in the absence of the logarithmic term. But this term is just the correct $O(\eta)$-term of the $\eta$-expansion of the prefactor $(i \tau)^{- i \eta}$ of the exact solutions (\ref{Gl87}). This prefactor is, up to a constant, given by $e^{i \omega(t) \, t}$ with asymptotically vanishing frequency $\omega(t) = |\eta| \ln|2 k_3 t|/t$. Clearly, if $\delta = 1/2$, the approximate Weyl TEO result can not restore this factor.

We now extend the above considerations to the case $n \geq 3$. It is convenient to introduce the quantity

\begin{equation}
\label{GlA02}
\Sigma_{l; n} = \sum_{m = l}^{n} (- 1)^{m + l} \sigma_m 
\end{equation}
which satisfies $\sigma_l \geq \Sigma_{l; n} \geq \sigma_l - \sigma_{l + 1}$, following from the domain of integration in (\ref{Gl37}), (\ref{Gl42}). This is given by the n-volume $\mathcal{V}_n(\sigma_1, ... ,\sigma_n)$ defined by $1 \geq \sigma_1  ... \geq \sigma_n \geq \sigma_A $. Due to the oscillating term exp$[- i \, \Sigma_{1;n} \, \tau]$ in (\ref{Gl37}), (\ref{Gl42}), only the integration over those subsets of $\mathcal{V}_n$ is of importance at large $\tau(s)$, which satisfy $\Delta \Sigma_{1;n} \ll 1$. For example, for the subsets $\mathcal{V}_n|_{\sigma_n \, \geq \, 1 - \varepsilon}$ or $\mathcal{V}_n|_{\mathfrak{b(s)} \, \geq \, \sigma_1}$ with

\begin{equation}
\label{GlA021}
\mathfrak{b}(s):= \sigma_A(s) + \varepsilon,   \ 0 < \varepsilon \ll 1  ,
\end{equation}
this condition is fulfilled. When integrating over $\mathcal{V}_n|_{\sigma_n \, \geq \, 1 - \varepsilon}$, approximation (\ref{Gl42}) is excellent since $f_1 \approx f_2$. The integration over $\mathcal{V}_n|_{\mathfrak{b} \,  \geq \, \sigma_1}$ is more complicated. In fact, the latter case is most critical w.r.t. the above approximation, because $f_1(\sigma_l)$ tends to infinity if $\sigma_l \rightarrow 0$, while $f_2(\sigma_l)$ stays finite. But the dominant contributions to the exact and approximate TEO (\ref{Gl36}) are in this case given by $I_0 \equiv  \widetilde{I}_0$ and $I_1 \equiv \widetilde{I}_1$, and we get only small corrections from higher terms $I_n, \widetilde{I}_n \, (n \geq 2)$. These are in the exact case at least of order $\eta_\delta \, \varepsilon^{2 \delta}$ and in the approximate case at least of order $\varepsilon^{2 \delta}$. 
Both cases, however, differ in that there are no  O($\eta_\delta$)-corrections for $\widetilde{I}_0$ if $\delta \neq 1/2$, see (\ref{GlA010}).

The integration over $\mathcal{V}_n  \,  (n > 1$) in the expressions for $\widetilde{I}_n$, $I_n$ is now rewritten so that all integrals over $\mathcal{V}_n|_{\mathfrak{b} \,  \geq \, \sigma_1}, \,\mathcal{V}_{n - 1}|_{\mathfrak{b} \,  \geq \, \sigma_2}, ...$ are split off

\begin{equation}
\label{GlA03}
\int\limits_{\mathcal{V}_n} [d \sigma]^n...=  \int\limits_{\mathcal{V}_n|_{\mathfrak{b}  \geq \sigma_1}} [d \sigma]^n...
+  \int\limits_{\mathfrak{b}}^{1} d\sigma_1 \int\limits_{\mathcal{V}_{n - 1}|_{\mathfrak{b}  \geq \sigma_2}} [d \sigma]^{n - 1}...  +  \int\limits_{\mathfrak{b}}^{1} d\sigma_1 \int\limits_{\mathfrak{b}}^{\sigma_1} d\sigma_2 \int\limits_{\mathcal{V}_{n - 2}|_{\sigma_2  \geq \sigma_3}} [d \sigma]^{n - 2}...    \nonumber
\end{equation}
The contributions of the split off integrals to the TEO (\ref{Gl36}) can be neglected by repeatedly applying an analogous argumentation as above. Hence, it suffices to consider instead of (\ref{Gl37}) and  (\ref{Gl42}), resp., the multiple integral

\begin{equation}
\label{GlA04}
\int\limits_{\mathfrak{b}}^{1}d\sigma_1 f_R(\sigma_1) e^{- i \sigma_1 \tau} ... \int\limits_{\mathfrak{b}}^{\sigma_{n - 2}}d\sigma_{n - 1} f_R(\sigma_{n - 1}) e^{i (-1)^{n - 1} \sigma_{n - 1} \tau} \int\limits_{\sigma_A}^{\sigma_{n - 1}}d\sigma_n f_1(\sigma_n) e^{i (-1)^n  \sigma_n \tau}  ,
\end{equation}
($R = 1, 2$). The asymptotic expansion of the exact n-th integral reads

\begin{equation}
\label{GlA05}
\begin{aligned}
\int\limits_{\sigma_A}^{\sigma_{n - 1}}d\sigma_n f_1(\sigma_n) e^{i (-1)^n  \sigma_n \tau} \ =& \ \frac{\Gamma(\delta) \, e^{[\mathrm{sign} (-1)^n \tau] i \pi \delta}}{[i (-1)^n \tau]^\delta} \, -  \,
\frac{\tau_A^\delta}{\delta} \frac{F_\delta(i (-1)^n \tau_A)}{\tau^\delta}
\\ 
&+ \, \frac{ f_1(\sigma_{n - 1}) \, e^{i (-1)^n \sigma_{n - 1} \tau}}{i (-1)^n \tau} \, + \, O\left( \frac{1}{\tau^2} \right)   ,
\end{aligned}
\end{equation}
($\tau_A \equiv 2 k_3 s_A/\mu$). Note that the first two terms are proportional to $\tau^{- \delta}$ and independent of $\sigma_{n - 1}$. These terms would not appear if $f_2$  had been used instead of $f_1$ on the l.h.s. of (\ref{GlA05}). It is now crucial that for any $\varepsilon > 0$ holds
\begin{equation}
\label{GlA06}
\int\limits_{\mathfrak{b}}^{\sigma_{l - 1}}d\sigma_l [f_R(\sigma_l)]^m e^{i (-1)^l \sigma_l \tau} \, = \, \frac{[f_R(\sigma_{l -1})]^m  e^{i (-1)^l \sigma_{l - 1} \tau} \, - \, [f_R(\mathfrak{b})]^m  e^{i (-1)^l \mathfrak{b} \tau }}{i (-1)^l  \tau \ [1 + O(\tau^{-1})]}   ,
\end{equation}
$ m \in \mathbb{N}$. Thus there are in (\ref{GlA06}), in contrast to (\ref{GlA05}), no more terms $\propto \tau^{- \delta}$. Hence, for nonzero $\varepsilon$ the exact ($R = 1$) and the approximate integral ($R = 2$) exhibit the very same asymptotic behavior. Since arbitrary powers of $f_R$ are again given by these functions (with suitable rescaling of $\delta$), and since the same is up to constant factors also true for integrals of arbitrary powers of $f_R$, it holds with (\ref{GlA05}) and (\ref{GlA06}) that expression (\ref{GlA04}) shows in both cases to leading order the same asymptotics. Explicitly calculating (\ref{GlA04}) for $n = 0, 2, 4, ...$ with the help of  (\ref{GlA05}), (\ref{GlA06}), and substituting these results into (\ref{Gl36}) yields in the exact case, $R = 1$, for the diagonal entries of the TEO

\begin{equation}
\label{GlA007}
\begin{aligned}
I_0 - I_2 + I_4 - ... \,  \sim &\ 1 + i \omega_1 t  + (i \omega_1 t)^2/2!  \, +  \, .... \,  -  \,  \mathcal{R}_\textbf{k}^2  \,\Gamma(\delta) [1 + O(\tau_A^\delta \, F_\delta(i \tau_A) )] 
\\
&\times \,  (1 - i \omega_1 t + ...) \, (- i \tau)^{\delta - 1} \, e^{- i \tau} \,  +  \,   O(\mathcal{R}_\textbf{k}^6)
\end{aligned}    \nonumber
\end{equation}
with real $\omega_1$, where

\begin{equation}
\label{GlA008}
\omega_1[\tau(t)] \, t \, = \, - i  \,  \mathcal{R}_\textbf{k}^2  \, \frac{1 - (\sigma_A + \varepsilon)^{2 \delta - 1}}{2 \delta - 1} \, (- i \tau)^{2 \delta - 1}   .
\end{equation}
Recall that $\mathcal{H}^2_\textbf{k} \equiv  -1_2$, and due to (\ref{GlA123}): $\mathcal{R}_\textbf{k}^2 = O(\eta_\delta)$. A similar result holds for the off-diagonal entries $k_+ [I_1 - I_3 + I_5 - ....]/\kappa$. If $\delta \to 1/2$, we recover from (\ref{GlA008}) the correct logarithmic behavior ($\varepsilon \equiv 0$). In the approximate case, $R = 2$, we obtain an analogous result with $\omega_2[\tau(t)] \, t  \propto$ sign$k_3 \, \eta_\delta |\tau|^{2 \delta - 1}$, i.e. $\omega_2   \propto  \omega_1$ if $\delta \neq 1/2$. But when $\delta \to 1/2$, then $\omega_2 t \to$ const., and again it turns out that in this special case the approximate TEO does not correctly reproduce a phase factor with vanishing frequency.

The above considerations imply that the approximate TEO agrees asymptotically with the exact one. As an aside, they also support the assumption that the summation of the $I_n$ according to  (\ref{Gl36}) must result in a phase of the form $e^{i \omega_1 t}$. While we could not prove this for the exact TEO, the feasible summation of the terms $\widetilde{I_n}$ leading to eq.s (\ref{Gl54}) indeed yields the corresponding factor $e^{i \omega_2 t}$ in the asymptotic limit at least for $0 < \delta \leq 1/2$.\footnote{See eq.s (\ref{GlB5}), (\ref{GlB9}) and (\ref{GlB12}). Note that from (\ref{GlB10}) follows: $\tau E_j(\tau) \propto \mathrm{sign}k_3 \,\eta_\delta |\tau|^{2 \delta - 1}  \propto \omega_R(\tau[t]) \, t$ $(j, R = 1, 2)$.}
\\
\renewcommand{\theequation}{B\arabic{equation}}
\setcounter{equation}{0}
\section*{Appendix B. Asymptotic expansion of the approximate Weyl TEO} 
Some effort is required to obtain analytic asymptotic expressions for eq.s (\ref{Gl54}), (\ref{Gl55}). We begin with rewriting $Z$

\begin{equation}
\label{GlB1}
Z(z; s) = \frac{2 \lambda(s)}{\mathcal{D} \, X(z; s)} \,  Z_+(z; s) - \widetilde{Z_+}(z; s) + \widetilde{Z_-}(z; s)
\end{equation}
where

\begin{equation}
Z_\pm (z;s) :=  J_{\pm \lambda}(\mathcal{D} X)  J_{\mp \lambda}(\mathcal{D} x) ,
\  \
\widetilde{Z_\pm} (z;s) :=  J_{1\pm  \lambda}(\mathcal{D} X)  J_{\mp \lambda}(\mathcal{D} x) .
\end{equation}
Here, $x, X, \lambda, Z, \mathcal{D}$ are defined in (\ref{Gl52}), (\ref{Gl55}), (\ref{Gl60}), and $k_3 \neq 0, \, \delta > 0$. Since a direct computation of the asymptotic behavior of Bessel's function $J_\lambda(x)$ becomes rather complicated when both quantities, order $\lambda$ and variable $x$, increase indefinitely, we employ instead \cite{Erdelyi}

\begin{equation}
\label{GlB31}
J_{\chi}(\alpha y) J_{\xi}(\beta y) = \frac{(2 \alpha)^{\chi} (2 \beta)^{\xi}}{\pi} \int \limits_{- \pi /2}^{\pi /2} d\theta \, e^{i \theta (\chi - \xi)} \, \left( \frac{\cos \theta}{\Lambda(\theta)} \right)^{\chi + \xi} \, J_{\chi + \xi}[\Lambda(\theta) y]   ,
\end{equation}
valid for Re$(\chi  + \xi) > - 1$,\footnote{This condition is not satisfied by the defining expression for $Z$, eq.  (\ref{Gl55}), but  by  (\ref{GlB1}).} where $\Lambda(\theta) := \sqrt{2 \cos \theta \, (\alpha^2 e^{i \theta} + \beta^2 e^{- i \theta})}$. With
 
\begin{equation}
\mathfrak{a}(s) := (1 - \delta)/\tau(s)
\end{equation}
($\tau$ as in (\ref{Gl411})),  $\alpha \equiv e^{(1 - \delta)z}$, $\beta \equiv 1$, and defining:

\begin{equation}
\mathcal{J}_n^{\pm}(v; z) = \frac{J_n[\Lambda(\frac{\pi}{2} \pm \mathfrak{a} v; z)\, |\mathcal{D} x| ]}{[ \Lambda(\frac{\pi}{2}  \pm \mathfrak{a} v; z) |\mathcal{D} x| ]^n}   ,  \nonumber
\end{equation}
one obtains:

\begin{equation}
\label{GlB4}
\begin{aligned}
Z_\pm(z; s) & =  \frac{\mp i \mathfrak{a}}{\pi} \, e^{\pm (1 - \delta) \lambda z} \, e^{\frac{\pi}{2 |\mathfrak{a}|}}  \int \limits_0^{\pi /2 |\mathfrak{a}|} dv \ \frac{\mathcal{J}_0^{\pm}(v; z)}{e^{(1 - i \mathfrak{a}) v}}    \left(1 + O(e^{- \frac{\pi}{2 |\mathfrak{a}|}})  \right)
\\
\widetilde{Z_\pm}(z; s)  = &\frac{ 2\mathfrak{a} \mathcal{D} X}{\pi } \, e^{\pm (1 - \delta) \lambda z} \, e^{\frac{\pi}{2 |\mathfrak{a}|}}  \int \limits_0^{\pi /2 |\mathfrak{a}|} dv \ 
\frac{\sin(\mp \mathfrak{a} v)}{ e^{\mp i \mathfrak{a} v}}  \  
\frac{\mathcal{J}_1^{\pm}(v; z)}{ e^{(1 - i \mathfrak{a})v}}
\left(1 + O(e^{- \frac{\pi}{2 |\mathfrak{a}|} } )   \right)   .
\end{aligned}
\end{equation}
For $|\tau| \gg 1$, the upper integration limits can be shifted to infinity. In this case it is also sensible to expand

\begin{equation}
\label{GlB04}
\Lambda(\pi/2 \mp \mathfrak{a} v; z)\, |\mathcal{D} x| = 2 B_{\pm} \sqrt{v} \, \left\lbrace 1 \mp i \coth[(1 - \delta) z] \, \mathfrak{a} v/2 + O[(\mathfrak{a} v)^2]   \right\rbrace       
\end{equation}
with

\begin{equation}
\label{GlB5}
B_{\pm}(z;s) := \sqrt{\pm \, i \mathfrak{a}(s) [e^{2(1 - \delta) z} - 1]/2 }  \ |\mathcal{D} x(s)|
\end{equation}
where $B^2_{\pm}  \propto  \pm i \, \mathrm{sign}k_3 \,\eta_\delta |\tau|^{2 \delta - 1}$. By means of the expression \cite{Erdelyi}:

\begin{equation}
J_{\alpha}(\beta y) = \beta^{\alpha} \sum_{n = 0}^{\infty} [y (1 - \beta)^2/2]^n J_{\alpha + n}(y)/n!    ,  \nonumber
\end{equation}
and with the restriction $\delta \leq 1/2$, we get:

\begin{equation}
\label{GlB5a}
\int \limits_0^{\infty} dv \, \frac{\mathcal{J}_0^{+}(v; z)}{e^{(1 - i \mathfrak{a}) v}} 
 \, \sim \, \frac{e^{- \frac{B^2_-}{1 - i\mathfrak{a}} } }{1 - i \mathfrak{a}} \,  \left\lbrace 1 
- \, \frac{ 2i \, \mathfrak{a} B^2_-}{\tanh(\mathfrak{d} z)} \, 
\frac{ 1 - \frac{1}{2}  \frac{B^2_-}{1 - i\mathfrak{a}}}{(1 - i \mathfrak{a})^2 } \, +  ... \right\rbrace   ,
\end{equation}
where $\mathfrak{d} := 1 - \delta$. Similar expansions are found for the remaining integrals in $Z_-, \widetilde{Z_\pm}$. Substituting these results for the integrals in (\ref{GlB4}) leads by use of  (\ref{GlB1}) and  (\ref{Gl55a}) to

\begin{equation}
\label{GlB8}
\int\limits_0^{1 - \sigma_A(s)} dz \, \left( \frac{e^{\lambda z}}{1 - z} \right)^\mathfrak{d} \, \frac{Z(z; x)}{Z(0; x)} \, \sim \, \sum\limits_{j=1}^{3} \int\limits_0^{1 - \sigma_A(s)} dz \, \frac{e^{\lambda \mathfrak{d}  z}}{(1 - z)^{\mathfrak{d}} } \, \mathcal{L}_j(z;s)   ,
\end{equation}
where

\begin{equation}
\label{GlB7}
\begin{aligned}
{\cal{L}}_1(z; s) :=&  \ e^{- \lambda^{\ast} \, \mathfrak{d} z} \, e^{\frac{- B^2_{-}}{1 - i\mathfrak{a}} }  \, 
\left\lbrace 1 - L_1 -  L_2 - ...   \right\rbrace   , 
\\
{\cal{L}}_2(z; s) :=& \ \frac{e^{ \lambda \mathfrak{d} z} \, [ \,  e^{\frac{- B^2_{-}}{1 - i\mathfrak{a}}} \, \left\lbrace 1 -  L_2 - ...\right\rbrace \ - \ e^{\frac{- B^2_{-}}{1 - 3i \mathfrak{a}}} \, \left\lbrace 1 -  L_3 - ...\right\rbrace ]}{2 \sinh(\mathfrak{d} z)}   ,
\\
{\cal{L}}_3(z; s) :=&\ \frac{e^{- \lambda \mathfrak{d} z} \, [ \ e^{\frac{- B^2_{+}}{1 - i\mathfrak{a}}} \, \left\lbrace 1 +  L_2 + ...\right\rbrace - e^{\frac{- B^2_{+}}{1 + i\mathfrak{a}}} \, \left\lbrace 1 -  L_2^{\ast} + ...\right\rbrace ]}{2 \sinh(\mathfrak{d} z)}   ,
\end{aligned}    \nonumber
\end{equation}
and

\begin{equation}
L_1(z; s) \, =  \frac{2 i \mathfrak{a}(s) \, B^2_-(z; s)}{(1 - i \mathfrak{a})^2 \tanh(\mathfrak{d} z)}  ,
\ \  \  \
L_2(z; s) \, = \, \frac{- i \mathfrak{a}(s) \, B^4_-(z; s)}{(1 - i \mathfrak{a})^3 \tanh(\mathfrak{d} z)}  ,
\nonumber
\end{equation}
and $L_3  := [(1 - i  \mathfrak{a})/(1 - 3i  \mathfrak{a})]^3  L_2$. It is now straightforward but tedious to evaluate the r.h.s. of (\ref{GlB8}), and one eventually finds with (\ref{Gl54})

\begin{equation}
\label{GlB9}
\begin{aligned}
\Big(K&_{\textbf{k}}^{(-)}(t|t_A) \Big)_{12}  \, \sim \,  \frac{k_+}{\kappa}  \frac{\sqrt{2 \eta_\delta}}{\mu} \left( \frac{\mu}{2} \right)^\delta  |\tau|^\delta \,
\Bigg[ \ \frac{e^{i \mathcal{D}^2 \tau   E_1(\tau)}}{\delta}  \left\lbrace 1 + O\left(\frac{\eta_\delta \mathcal{D}^2}{|\tau|^{2 \mathfrak{d}}} \right)  \right\rbrace
\\
& \times \left\lbrace F_\delta(- i \tau [1 + \mathcal{D}^2  E_2(\tau)])  -  \left( \frac{\tau_A}{\tau} \right)^\delta  F_\delta(- i \tau_A [1 + \mathcal{D}^2  E_2(\tau)]) \right\rbrace
\\
& + \frac{e^\mathfrak{d} \, \eta_\delta  \, \mathcal{D}^2}{2 \delta \, (\mu/2)^{2 \mathfrak{d}}} \, \frac{e^{- i \tau} }{|\tau|^{2 \mathfrak{d}} }   \, \left\lbrace 1 + O\left( \frac{\eta_\delta \mathcal{D}^2}{|\tau|^{2 \mathfrak{d} - 1}} \right) \right\rbrace 
\, \left\lbrace F_\delta(- \mathfrak{d}) -  \frac{\tau^\delta_A}{\tau^\delta}  F_\delta\left(- \mathfrak{d} \frac{\tau_A}{\tau}\right) \right\rbrace
\\
& + \frac{e^{2 \mathfrak{d}} \, \eta_\delta \, \mathcal{D}^2}{2 \delta  \, (\mu/2)^{2 \mathfrak{d}}}   \, 
\frac{1 + O\left( \frac{\eta_\delta \mathcal{D}^2}{|\tau|^{2 \mathfrak{d} - 1}} \right)}{|\tau|^{2 \mathfrak{d}}} \,  
 \left\lbrace F_\delta(- 2\mathfrak{d} - i \tau) - \frac{\tau^\delta_A}{\tau^\delta}  F_\delta \left(- 2\mathfrak{d} \frac{\tau_A}{\tau} - i \tau_A \right)  \right\rbrace  \Bigg]
\end{aligned}
\end{equation}
where $F_\alpha(x) := \, _1F_1(\alpha; \alpha + 1; x)$, and

\begin{equation}
\label{GlB10}
E_1(\tau) = \eta_\delta \, \frac{e^{2 \mathfrak{d}} - 1}{4 \mathfrak{d}} \, 
\frac{ \left( \frac{\mu}{2} |\tau| \right)^{- 2 \mathfrak{d}} }{1 - i \mathfrak{d}/\tau},
\  \  \
E_2(\tau) =  \, \frac{\mathfrak{d} e^\mathfrak{d}}{\sinh (\mathfrak{d})} \, E_1(\tau)  .
\end{equation}
The determination of the asymptotic limit of $K^{(-)}_{11}$ proceeds analogously. With $R :=  Z_+   +  Z_-$ and definition

\begin{equation}
\label{GlB13}
G(\tau) =   \, \tau \,  E_1(\tau)/(e^{2 \mathfrak{d}} - 1)   ,
\end{equation}
the same manipulations as above lead to

\begin{equation}
\label{GlB12}
\begin{aligned}
\Big( K&_{\textbf{k}}^{(-)}(t|t_A) \Big)_{11} \, \sim \,  1 + 2 i \mathcal{D} \mathfrak{d}  \, G(\tau) \Bigg[ \ \frac{e^\mathfrak{d} }{\delta}  \left\lbrace 1 - i \mathcal{D}^2 G(\tau)  \right\rbrace  
\Big\lbrace F_\delta(-\mathfrak{d}) -  \frac{\tau_A^\delta}{\tau^\delta}
\\
& \times F_\delta\left(-\mathfrak{d} \, \frac{\tau_A}{\tau}\right) \Big\rbrace
 + i \, \frac{e^{3 \mathfrak{d}}  \mathcal{D}^2}{\delta} \, G(\tau) \left\lbrace F_\delta(-3 \mathfrak{d}) - \frac{\tau_A^\delta}{\tau^\delta}  F_\delta \left(-3 \mathfrak{d} \frac{\tau_A}{\tau} \right) \right\rbrace
\\
& + O(\mathcal{D}^4 G^2(\tau))  - \frac{e^{- i \tau (1 + \mathcal{D}^2  E_1(\tau) )}}{\delta}  \left\lbrace 1 + 
O\left(\frac{\eta_\delta \mathcal{D}^2}{|\tau|^{2 \mathfrak{d}}} \right) \right\rbrace
\\
& \times \left\lbrace F_\delta( i \tau [1 + \mathcal{D}^2  E_2(\tau) ]) - \frac{\tau_A^\delta}{\tau^\delta} F_\delta( i \tau_A [1 +  \mathcal{D}^2 E_2(\tau)]) \right\rbrace \Bigg]   ,
\end{aligned}
\end{equation}
If $\delta = 1/2$, one gets ($\eta \equiv  \eta_{\frac{1}{2}} \, \mathrm{sign}k_3  \equiv  \kappa^2/2k_3)$

\begin{equation}
\label{GlB14}
\begin{aligned}
\left( K_{\textbf{k}}^{(-)}(t|t_A) \right)_{11} & \sim \, 1 + i \, \mathcal{D}(1/2) \, \frac{\eta}{\mu} \ \Big[2 \, e^{1/2}  \{1 + O(\eta) \} 
\\
&\times \, \left\lbrace F_{1/2}(- 1/2) + O(\sqrt{\tau_A/\tau}) \right\rbrace  \, +  \, O(e^{- i \tau}/\sqrt{|\tau|}) \Big]   ,
\\
\left( K_{\textbf{k}}^{(-)}(t|t_A) \right)_{12}  & \sim \, \frac{k_+}{\kappa} \sqrt{ \frac{|\eta|}{\mu}} \,  \sqrt{|\tau|} \
\Big[ 2 \left\lbrace 1 + O(\eta) + O(\eta/\tau) \right\rbrace
\\
&\times \left\lbrace F_\delta(- i \tau [1 + O(\eta/\tau)])  -  O( \sqrt{\tau_A/\tau})  \right\rbrace  +  O(\eta/\tau)  \Big]   .
\end{aligned}
\end{equation}
\\
\renewcommand{\theequation}{C\arabic{equation}}
\setcounter{equation}{0}
\section*{Appendix C. PT as conformal map and the PT limiting case} 
The close relationship between the diffeomorphism $f^{(a)}$ defined in (\ref{Gl40a}) and the PT (\ref{Gl39}) can be seen as follows: Consider  a tangent vector $X_p \in  T_p(M)$ to $M$ at $p \in M$ and define the vector space homomorphism $f^{(a)}_{\ast p}: T_p(M) \rightarrow T_q(\widetilde{M})$ by $f^{(a)}_{\ast p}(X_p)(h) := X_p(h \circ f^{(a)})$. The points $p \in M$ and $q = f^{(a)}(p) \in \widetilde{M}$ are in local coordinates given by $(x_0, ... ,x_{n - 1})$ and $(\widetilde{x}_0, ... , \widetilde{x}_{n - 1})$, and $h \in C^\infty(\widetilde{M})$. $f^{(a)}_\ast$ is the differential or Jacobian of $f^{(a)}$. The dual map $f_q^{(a)^\ast}: T^\ast_q(\widetilde{M}) \rightarrow T^\ast_p(M)$ is defined by $(f_q^{(a)^\ast} \omega_q)(X_p) := \omega_q(f^{(a)}_{\ast p}(X_p))$, with $\omega_q \in T^\ast_q(\widetilde{M})$. Consider now $f^{(a)}$ as conformal map satisfying

\begin{equation}
\label{Gl40b}
f_q^{(a)^\ast} \, \boldsymbol{\widetilde{g}}|_q  \, = \, \Lambda^2 \, \boldsymbol{g}|_p  ,
\end{equation}
where $\Lambda(x) \equiv a x_0^{a - 1}$, and $\boldsymbol{\widetilde{g}} \in T^\ast (\widetilde{M}) \otimes T^\ast (\widetilde{M})$ and $\boldsymbol{g}  \in T^\ast (M) \otimes T^\ast (M)$ are locally represented by $\boldsymbol{\widetilde{g}}|_q = \widetilde{g}_{\mu \nu} (\widetilde{x}) \, d\widetilde{x}^\mu \otimes d\widetilde{x}^\nu$ and $\boldsymbol{g}|_p = g_{\mu \nu}(x) \, dx^\mu \otimes dx^\nu$. Since with basis vectors $\{\partial_\alpha\}_\alpha \in T_p(M)$ holds: ($f_q^{(a)^\ast} \, \boldsymbol{\widetilde{g}}|_q) (\partial_\alpha, \partial_\beta) = \boldsymbol{\widetilde{g}}|_q(f^{(a)}_{\ast p}(\partial_\alpha), f^{(a)}_{\ast p}(\partial_\beta))$, eq. (\ref{Gl40b}) assumes in a coordinate neighborhood the form

\begin{equation}
\label{Gl40c}
\widetilde{g}_{\alpha \beta}(\widetilde{x}) \ \frac{\partial (\widetilde{x}^{\alpha} \circ f^{(a)}) }{\partial x^\nu} \, \frac{(\widetilde{x}^{\beta} \circ f^{(a)})}{\partial x^\mu} \, = \, \Lambda^2(x) \, g_{\mu \nu}(x)   .
\end{equation}
$\widetilde{x}^{\alpha}$ denotes the $\alpha$-th coordinate function $\widetilde{U}_q \rightarrow \mathbb{R}$, with $\widetilde{x}^{\alpha} \circ f^{(a)}(p) \equiv \widetilde{x}^{\alpha}(q) = \widetilde{x}_{\alpha}  \ (\widetilde{U}_q \subset \widetilde{M})$. Then one obtains with (\ref{Gl40}), (\ref{Gl40a}), (\ref{Gl40c}):

\begin{equation}
\widetilde{g}_{\alpha \beta}(\widetilde{x}) = \mathrm{diag} \, (1, - \widetilde{x}_0^{2 \widetilde{\nu}}, - \widetilde{x}_0^{2 \widetilde{\nu}}, - \widetilde{x}_0^{2 - 2 \widetilde{\mu}})     ,  \nonumber
\end{equation}
where

\begin{equation}
\label{Gl40d}
\widetilde{\nu} = 1 - a^{- 1} (1 - \nu), \ \ \widetilde{\mu} = a^{- 1} \mu   .
\end{equation}
Eq. (\ref{Gl40d}) actually defines the PT inverse to (\ref{Gl39}), a consequence of $f^{(a)^\ast}$ having the "opposite direction" compared to $f^{(a)}$. As a result we get: $\boldsymbol{\widetilde{g}} \equiv \boldsymbol{{g}^{'}}$. Thus, (\ref{Gl40a}) together with constraint (\ref{Gl40aa}) is for massless Dirac-spinors equivalent to (\ref{Gl39}) together with normalization condition (\ref{Gl26a}).\footnote{(\ref{Gl26a}) enforces the correct transformation of the normalization constants $c^{(j, \mp)}_\textbf{k}$ of the spinors (\ref{Gl22}), when the PT is applied, which is automatically taken care of by $f^{(a)}$. Of course, (\ref{Gl39}) and (\ref{Gl40a}) are not equivalent, neither are (\ref{Gl26a}) and (\ref{Gl40aa}).}

As has already been mentioned, the PT or the diffeomorphism $f^{(a)}$ generate all solutions of a given equivalence class of exact massless spinor solutions, with the only prerequisite being the knowledge of a single arbitrary exact solution of this class. Moreover, according to \cite{Wollensak1} it should be a general feature of every such equivalence class, that in the limiting case $a \rightarrow 0$ its solutions must always approach those solutions of the Weyl-Dirac equation where the background spacetime is given by the special fFLRW line element (\ref{Gl27}) with $\nu = 1, \, \mu = 0$. We prove that $all$ exact Weyl TEOs (\ref{Gl35}) - (\ref{Gl37}) satisfy this criterion, and in fact also $all$ related approximate Weyl TEOs given by eq.s (\ref{Gl53}) - (\ref{Gl55}). We begin with the exact case (\ref{Gl36}), (\ref{Gl37}) and apply (\ref{Gl40a}), (\ref{Gl40aa}), that is we consider $I_n(s^a)$ with $\textbf{k} \to \textbf{k}^{'} := \textbf{k}/a$ and $\tau \to \tau^{'} := 2 k_3 s^a/a \mu \ (a, \sigma_A > 0)$

\begin{equation}
\label{GlC1}
I_n(s^a)  = \left(  \frac{\kappa s^{a \delta}}{a \mu} \right)^n  \int\limits_{\sigma^a_A}^1 d\sigma_1 \int\limits_{\sigma^a_A}^{\sigma_1} d\sigma_2 ...\int\limits_{\sigma^a_A}^{\sigma_{n-1}}    d\sigma_n 
\prod\limits_{l = 1}^{n} \sigma_l^{\delta -1} 
e^{i (- 1)^l \sigma_l \, \tau^{'}}   .   \nonumber
\end{equation}
If $a \ll 1$, then: $\sigma_j = 1 - \epsilon_j \ (\epsilon_j \ll 1, \, j = 1, ..., n)$, which means that for $a \to 0$ approximation (\ref{Gl41}) becomes exact. Hence, exact (eq.s (\ref{Gl36}), (\ref{Gl37})) and approximate TEO expressions (eq.s (\ref{Gl36}), (\ref{Gl42}))  tend to the $same$ outcome when $a \to 0$. It therefore suffices to consider the approximate TEO given by eq.s (\ref{Gl53}) - (\ref{Gl55}), but here necessarily with $\mathcal{D} \equiv 1$, because (\ref{Gl60}) with $\mathcal{D} \neq 1$ spoils the PT limiting case ($\mathcal{D}(\delta)$ is invariant under the PT!).

Rewriting eq.s (\ref{Gl54}), (\ref{Gl55}) appropriately by means of (\ref{Gl40a}), (\ref{Gl40aa})
\begin{equation}
\label{GlC2}
\begin{aligned}
\left( K_{\textbf{k}/a}^{(-)}(t^a|t^a_A) \right)_{11} & = 1 - \frac{\kappa s^{a \delta}}{a \mu} \, \int\limits_0^{1 - \sigma^a_A} dz \, \left( \frac{e^{\lambda^{' \ast} z}}{1 - z} \right)^{1 - \delta} \, \frac{R(z; x^{'})}{Z(0; x^{'})}  \ ,
\\
\left( K_{\textbf{k}/a}^{(-)}(t^a|t^a_A) \right)_{12} & =  \frac{k_+ s^{a \delta}}{a \mu} \, e^{- i \tau^{'}} \int\limits_0^{1 - \sigma^a_A} dz  \, \left( \frac{e^{\lambda^{'} z}}{1 - z} \right)^{1 - \delta} \, \frac{Z(z; x^{'})}{Z(0; x^{'})}  ,
\end{aligned}
\end{equation}
where $x^{'} := \kappa s^{a \delta}/a \mu (1 - \delta), \, \lambda^{'} := 1/2 + i \tau^{'}/2(1 - \delta)$, and doing the same with eq.s  (\ref{Gl106a}) (substitute $1 - \delta$ for $\epsilon$) yields

\begin{equation}
\label{GlC3}
J_{\pm \lambda^{' \ast} + n}(x^{'}) \, J_{\pm \lambda^{'}}(x^{'})  \, = \,  \frac{1}{\pi} \int\limits_{- \frac{\pi}{2}}^{\frac{\pi}{2}} dv \, e^{\frac{\mp \tau^{'}  v}{1 - \delta}} \, e^{ - i n v} J_{n \pm 1}(2x^{'}  \cos v)   .
\end{equation}
Since $a \ll 1$, eq. (\ref{GlC3}) can be written as
\begin{equation}
\label{GlC4}
\begin{aligned}
\frac{\pi \tau^{'}  e^{\frac{- \pi |\tau^{'}|}{2 (1 - \delta)}} }{1 - \delta}  \,  J_{\pm \lambda^{' \ast} + n}(x^{'}) \, J_{\pm \lambda^{'}}(x^{'})  \, = \  &\frac{(\pm i \,  \mathrm{sign} k_3)^n}{1 - O(e^{- c \, \pi |\tau^{'}|})}
\, \int\limits_{0}^{\infty} dv \, \frac{J_{n \pm 1} \left(\frac{\kappa \, [1  \, +  \, O(a)]}{|k_3|}  \, v \right)}{e^{[1 \mp i  \, O(a)] v}} 
\\
& \stackrel{a \rightarrow 0}{\rightarrow} \ 
\frac{|k_3|}{k} \, (\pm i)^n  \, \left( \frac{k - |k_3|}{\kappa} \right)^{n \pm 1}   .
\end{aligned}
\end{equation}

It is convenient to introduce the new integration variable $\zeta := z/(1 - \sigma^a_A)$ with $\zeta \in [0; 1]$. Then, since $z = O(a)$

\begin{equation}
\label{GlC5}
\frac{\pi \tau^{'}  e^{\frac{- \pi |\tau^{'}|}{2 (1 - \delta)}} }{1 - \delta}  \,   Z(z; x{'}) \, \stackrel{a \rightarrow 0}{\rightarrow} \  \frac{\kappa |k_3|}{k} 
\left\lbrace \frac{e^{i  \,  \zeta \, k \, \mathrm{sign} k_3 \, \ln (t/t_A)}}{k  \mathrm{sign} k_3 - k_3}  \, +   \frac{e^{- i \, \zeta \, k \, \mathrm{sign} k_3 \, \ln (t/t_A)}}{k  \mathrm{sign} k_3 + k_3}   \right\rbrace
\end{equation}
where (\ref{Gl55}), (\ref{Gl106}), (\ref{Gl106a}), (\ref{GlC4}) have been utilized. Analogously one gets

\begin{equation}
\label{GlC6}
\frac{\pi \tau^{'} \ e^{- \frac{\pi |\tau^{'}|}{2 (1 - \delta)}}}{1 - \delta}   \,  R(z; x{'}) \, \stackrel{a \rightarrow 0}{\rightarrow} \  \frac{2 |k_3|}{k} \, \sin \left(\zeta \, k  \, \mathrm{sign} k_3  \, \ln \frac{t}{t_A}  +  O(a) \right)   ,
\end{equation}
and insertion of eq.s (\ref{GlC5}), (\ref{GlC6}) into (\ref{GlC2}) yields for $a \to 0$

\begin{equation}
\label{GlC7}
\begin{aligned}
\left( K_{\textbf{k}}^{(-)}(t|t_A) \right)_{11} \, &= \, e^{- i k_3 \ln (t/t_A)} \, \left[ \cos \left( k \ln \frac{t}{t_A} \right) \, + \, i  \, \frac{k_3}{k}  \, \sin \left( k \ln \frac{t}{t_A} \right) \right]   ,
\\
\left( K_{\textbf{k}}^{(-)}(t|t_A) \right)_{12} \, &= \, \frac{k^{'}_+}{k} \, e^{- i k_3 \ln (t/t_A)}  \, \sin \left( k \ln \frac{t}{t_A} \right)    ,
\end{aligned}
\end{equation}
with $k^{'}_+ := (k_2 + i k_1) e^{2 i k_3 \ln (t_{\widetilde{A}}/t_A )}$. This is the exact Weyl TEO in the background spacetime $ds^2 = dt^2 - t^2 (dx^2 + dy^2 + dz^2)$, which is a special result of (\ref{Gl1202}) for $\delta = 1, \, \mu \to 0, \, \nu \equiv 1 -  \mu  \to 1$ (recall that $s \equiv t^\mu$). While the limiting case of conformal flatness leads to the entire class of exact spinor solutions, the PT limiting case reduces only (as it should) to that special solution, which corresponds to the PT fixed point $ \mu = 0, \, \nu = 1$ \cite{Wollensak1}.

\end{appendix}


\begin{thebibliography}{50}

\bibitem{Brill} D. R. Brill and J. A. Wheeler, Interaction of Neutrinos and Gravitational Fields, $Rev. \, Mod. \, Phys.$ \textbf{29} (1957), 465 - 479.

\bibitem{Parker} L. Parker, Quantized Fields and Particle Creation in Expanding Universes. II, $Phys. \, Rev. \, D$ \textbf{3} (1971), 346 - 356.

\bibitem{FLRW-Quant} M.A. Castagnino, L. P. Chimento, D.D. Harari and C.A. Nunez, A spin-$\frac{1}{2}$ particle formalism in curved space-time, $J. \, Math. \, Phys.$ \textbf{25} (1984), 360 - 367.
L. P. Chimento and M. S. Mollerach, Quantum vacuum definition for spin-$\frac{1}{2}$ fields in Robertson-Walker metrics, $Phys. \, Rev. \, D$ \textbf{34} (1986), 3689 - 3697.
E. Montaldi and A. Zecca, Second Quantization of the Dirac Field: Normal Modes in the Robertson-Walker Space-Time, $Int. \, J. \, Theor. \, Phys.$ \textbf{37} (1998), 995 - 1009.

\bibitem{Barut} A. O. Barut and I. H. Duru, Exact solutions of the Dirac equation in spatially flat Robertson-Walker space-times, $Phys. \, Rev. \, D$ \textbf{36} (1987), 3705 - 3711.

\bibitem{Cotaescu} I. I. Cotaescu, Polarized Dirac fermions in de Sitter spacetime, $Phys. \, Rev. \, D$ \textbf{65} (2002), 084008 (1 - 9); arXiv:hep-th/0109199.

\bibitem{Candelas1} P. Candelas and D. J. Raine, General-relativistic quantum field theory: An exactly soluble model, $Phys. \, Rev. \, D$ \textbf{12} (1975), 965 - 974.

\bibitem{Koksma} J. F. Koksma and T. Prokopec, The fermion propagator in cosmological spaces with constant deceleration, $Class. \, Quant. \, Grav.$ \textbf{26} (2009), 125003 (1 - 28); arXiv:0901.4674 [gr-qc].

\bibitem{Zel`dovich2} Ya. B. Zel`dovich, Particle production in cosmology, $JETP \, Lett.$ \textbf{12} (1970), 307 - 311.

\bibitem{Hu} B. Hu and L. Parker, Anisotropy damping through quantum effects in the early universe, $Phys. \,Rev. \, D$ \textbf{4} (1978), 933 - 945.

\bibitem{BKL} V. A. Belinskii, I. M. Khalatnikov, and E. M. Lifshitz, Oscillatory approach to a singular point in the relativistic cosmology, $Adv. \, Phys.$ \textbf{19} (1970), 525 - 573.

\bibitem{Misner} C. W. Misner, Mixmaster Universe, $Phys. \, Rev. \, Lett.$ \textbf{22} (1969), 1071 - 1074.

\bibitem{Pitrou} C. Pitrou, T.S. Pereira, and J. P. Uzan, Predictions from an anisotropic inflationary era, $JCAP$ \textbf{04} (2008) 004(1 - 48); arXiv:0801.3596 [astro-ph].
H. C. Kim and M. Minamitsuji, Scalar field in the anisotropic universe, $Phys. \, Rev. \, D$ \textbf{81} (2010), 083517 (1 - 15); arXiv:1002.1361 [gr-qc].

\bibitem{Gm} A. E. G$\ddot{\mathrm{u}}$mr$\ddot{\mathrm{u}}$kc$\ddot{\mathrm{u}}$oglu, L. Kofman, and M. Peloso, Gravity waves signatures from anisotropic preinflation, $Phys. \, Rev. \, D$ \textbf{78} (2008), 103525 (1 - 21); arXiv:0807.1335 [astro-ph].

\bibitem{Armendariz} C. Armendariz-Picon, Could dark energy be vector-like?, $JCAP$ \textbf{07} (2004) 007 (1 - 22); arXiv:astro-ph/0405267.
 A. Golovnev, V. Mukhanov and V. Vanchurin, Vector inflation, JCAP \textbf{06} (2008) 009 (1 - 7); arXiv:0802.2068 [astro-ph].
 T. Koivisto and D. F. Mota, Vector field models of inflation and dark energy, $JCAP$ \textbf{08} (2008) 021 (1 - 24); arXiv:0805.4229 [astro-ph].

\bibitem{Malaknejad} A. Maleknejad, M. M. Sheikh-Jabbari and J. Soda, Gauge Fields and Inflation, $Phys. \, Rept.$ \textbf{528} (2013), 161 - 261; arXiv:1212.2921 [hep-ph].

\bibitem{Henneaux} M. Henneaux, Bianchi type-I cosmologies and spinor fields, $Phys. \, Rev. \, D$ \textbf{21} (1980), 857 - 863.

\bibitem{Saha} B. Saha and T. Boyadjiev, Bianchi type-I cosmology with scalar and spinor fields, $Phys. \, Rev. \, D$ \textbf{69} (2004), 124010 (1 - 12).

\bibitem{Zel`dovich1} Ya. B. Zel`dovich and A. A. Starobinski, Particle production and Vacuum Polarization in an Anisotropic Gravitational Field, $Sov. \, Phys. \, JETP$ \textbf{34} (1971), 1159 - 1163.

\bibitem{Birrell} N. D. Birrell and P. C. W. Davies, Massive particle production in anisotropic space-times, $J. \, Phys. \, A$ \textbf{13} (1980), 2109 -2120.
 
\bibitem{Lotze} K. H. Lotze, Production of massive spin-$\frac{1}{2}$ particles in anisotropic spacetimes,  $Class. \, Quantum \, Grav.$ \textbf{3} (1986), 81 - 95.

\bibitem{Wollensak1} M. Wollensak, Massless fermions in planar Bianchi-type-I universes: exact and approximate solutions, $Eur. \, Phys. \, J. \, C$ \textbf{81} (2021), 507 (1 - 11); arXiv:1909.12834 [hep-th]

\bibitem{Trautman} A. Trautman, in \textit{General Relativity and Gravitation}, edited by A. Held (Plenum Press, New York, 1980), p. 287.

\bibitem{Sexl} R. U. Sexl and H. K. Urbantke, \textit{Gravitation und Kosmologie}, 3rd ed. (BI-Wiss.-Verl., Mannheim, 1987).

\bibitem{Wollensak2}  M. Wollensak, Maxwell fields in anisotropic space-times, $J. \, Math. \, Phys.$ \textbf{39} (1998), 5934 -5945.

\bibitem{Lif} E. M. Lifshitz and I. M. Khalatnikov, Investigations in relativistic cosmology, $Adv. \, Phys.$ \textbf{12}(1963), 185 - 249 .

\bibitem{Weissinger} J. Weissinger, Zur Theorie und Anwendung des Iterationsverfahrens, $Math. \, Nachr.$ \textbf{8} (1952), 193 - 212.

\bibitem{Tsamis} N. C. Tsamis and R. P. Woodard, Plane waves in a general Robertson-Walker background,  $Class. \, Quantum \, Grav.$ \textbf{20}, 5205 - 5223 (2003); arXiv:astro-ph/0206010.
N. C. Tsamis and R. P. Woodard, General lane wave mode functions for scalar-driven cosmologies, $Class. \, Quantum \, Grav.$ \textbf{21}, 93 - 102 (2003); arXiv:astro-ph/0306602.

\bibitem{Abramowitz} M. Abramowitz and I. A. Stegun, \textit{Pocketbook of Mathematical Functions} (H. Deutsch, Frankfurt, 1984).

\bibitem{Erdelyi} A. Erdelyi (ed.), \textit{Higher Transcendental Functions} (McGraw - Hill, New York, 1955),  Vol. 2.

\bibitem{Kamke} E. Kamke, \textit{Differentialgleichungen, L$\ddot{o}$sungsmethoden und L$\ddot{o}$sungen}, 9th ed. (Teubner, Stuttgart, 1977)

\bibitem{Kramer} D. Kramer, H. Stephani, M. MacCallum and E. Herlt, \textit{Exact Solutions of Einstein`s Field Equations} (Deutscher Verlag d. Wiss., Berlin, 1980).

\bibitem{Pimentel} L. O. Pimentel, Weyl Equation in some Anisotropic Stiff Fluid Universes, $Int. \, J. \, Theor. \, Phys.$ \textbf{32} (1993), 979 - 984.

\bibitem{Ryzhik} I. S. Gradshteyn and I. M. Ryzhik, \textit{Table of Integrals, Series, and Products} (Academic Press, Orlando, 1980).

\bibitem{Chimento} L. P. Chimento and M. S. Mollerach, Dirac equation in Bianchi I metrics, $Phys. \, Lett. \, A$ \textbf{121} (1987), 7 - 10.

\bibitem{Castagnino} M. A. Castagnino, C. D. El Hasi, F. D. Mazzitelli, and J. P. Paz, On the Dirac equation in anisotropic backgrounds, $Phys. \, Lett. \, A$ \textbf{128} (1988), 25 - 28.

\bibitem{Slater} L. J. Slater, \textit{Generalized Hypergeometric Functions} (Cambridge Univ. Press, 1966)

\bibitem{Kim} S. K. Kim, An Asymtotic Expansion of a Hypergeometric Function $_2F_2(1, 2 \alpha; \rho_1, \rho_2 ; z)$ $Math. \, Comp.$ \textbf{26} (1972), 963.

\end{thebibliography}
\end{document}